# Near-field radiative heat transfer between irregularly shaped dielectric particles modeled with the discrete system Green's function method

Lindsay P. Walter[1], Eric J. Tervo[2,3,*], and Mathieu Francoeur[1,†]

[1]Radiative Energy Transfer Lab, Department of Mechanical Engineering, University of Utah, Salt Lake City, UT 84112, USA

[2]Department of Electrical and Computer Engineering, University of Wisconsin-Madison, Madison, WI 53706, USA

[3]Department of Mechanical Engineering, University of Wisconsin-Madison, Madison, WI 53706, USA

**ABSTRACT**

Near-field radiative heat transfer (NFRHT) between irregularly shaped dielectric particles made of $SiO_2$ and morphology characterized by Gaussian random spheres is studied. Particles are modeled using the discrete system Green's function (DSGF) approach, which is a volume integral numerical method based on fluctuational electrodynamics. This method is applicable to finite, three-dimensional objects, and all system interactions are defined independent of thermal excitation by a generalized system Green's function. The DSGF method is deemed suitable to model NFRHT between irregularly shaped particles after verification against the analytical solution for chains of two and three $SiO_2$ spheres. The NFRHT results reveal that geometric irregularity in particles leads to a reduction of the total conductance from that of comparable perfect spheres at vacuum separation distances smaller than the particle size, a regime in which NFRHT is a surface phenomenon. At vacuum separation distances larger than the particle size,

---

[*] Corresponding author. Email address: tervo@wisc.edu
[†] Corresponding author. Email address: mfrancoeur@mech.utah.edu

NFRHT becomes a volumetric process, and the total conductance between irregularly shaped particles converges to that of comparable perfect spheres. Spectral analysis reveals, however, that particle irregularity leads to damping and broadening of resonances at all separation distances, thereby highlighting the importance of the DSGF method for spectral engineering in the near field. The reduced spectral coherence when particle size is larger than the vacuum separation distance is attributed to coupling of surface phonon-polaritons within the randomly generated, distorted particle features. For particle size smaller than the vacuum separation distance, resonance broadening and damping is linked with the multiple localized surface phonon modes supported by the composite spherical harmonic morphologies of the Gaussian random spheres. This work has direct implications for thermal management of packed particle systems, with applications in radiative property control, electronics, energy conversion, and nanomanufacturing.

## I. INTRODUCTION

When objects are small or closely separated as compared to the characteristic thermal wavelength, Planck's classic theory of thermal radiation is no longer valid. This regime of small length scales is considered the thermal near field and is distinguished from the far-field regime by wave interference and the tunneling of evanescent electromagnetic modes [1–3] that can lead to radiation transport exceeding Planck's blackbody limit [4–14]. Surface phonon-polaritons (SPhPs), which are coupled transverse optical phonons and electromagnetic waves, provide additional pathways and tuning capability for near-field radiative heat transfer (NFRHT) [15]. The use of SPhPs to control NFRHT has received growing attention in recent years. Researchers have focused on harnessing SPhPs to increase the thermal conductivity of nanowires [16] and thin films [17–20]; on coupling SPhPs with other resonances, such as surface plasmon-polaritons [14,21–24], magnetoplasmon polaritons [25], and zone-folded longitudinal optical phonons [26], to control dispersion relations and manipulate heat flux; and on exploiting SPhPs for design of thermal circuit elements, such as switches [27] and diodes [28–30]. In each of these cases, the geometries and materials of micro/nanostructures are crucial in determining SPhP behavior and the resulting NFRHT.

Notably, it has been shown that NFRHT between micro/nanoscale particles is affected by geometric variation, especially around resonances [31]. While current NFRHT experimental techniques are limited to measurements of spheres with microscale radii and probe tips with radii of curvature greater than about 30 nm [6,32,33], models of heat exchange between particles below these cutoffs are important for developing a more complete understanding of particle near-field thermal interactions that can be applied to design superstructures with measurable heat exchanges. The need to understand NFRHT between particles arises in diverse fields, from energy conversion to biomedical applications to meteorology. Micro/nanoparticles are being investigated, for instance, to control radiative properties via metamaterials [34,35], to build biomimetic photovoltaic devices [36], and to treat cancer in photothermal ablation therapies [37–39]. Naturally occurring particles, such as dust and ash, are also gaining importance in meteorological models [40,41]. In these and related applications, the particles under consideration are usually irregularly shaped and contain geometric defects. To accurately account for NFRHT effects in such a diverse range of particle systems, it is therefore necessary that models support the complex three-dimensional structures emblematic of real-world particles.

Within the field of NFRHT, a few different particle-like structures have been studied that consider variations in particle arrangement and geometry. Researchers have modeled the effects of stretching and flattening of perfect spheroids [31,42], rotating one perfect cylinder [43,44] or spheroid [42,45] with respect to another, and varying dipole array structure [46–51]. In the above-mentioned studies, however, individual particles are modeled as geometrically regular and highly symmetric, with random geometric deviations relegated to the system level. As such, the effect of particle-level geometric irregularity on NFRHT has not been addressed and remains unknown.

A large part of why NFRHT between irregularly shaped particles remains underexplored is due to the difficulty in modeling three-dimensional objects of arbitrary shape. While analytical solutions exist for spheres [52–54] and point particles within the dipole limit [46,47,49,55,56], full-scale numerical models that can account for arbitrary geometries, such as finite difference time domain methods [57–62], boundary element methods [63–65], and volume integral approaches [66–72], are often computationally expensive. Of these methods, the boundary element method in the fluctuating-surface-current approach [43,44] is one of the most efficient techniques for modeling NFRHT between complex shaped objects. In this approach, computational efficiency is achieved through restriction of the solution space to the surface of thermal objects. If resolving the internal heat transfer physics of the thermal objects is desired, or if the thermal objects are characterized by nonuniform temperature and/or material, approaches other than the boundary element method are necessary, such as volume integral methods. However, thermal objects characterized by large dielectric functions, such as metals, are generally difficult to model with volume integral methods [67].

In this paper, we address the knowledge gap of NFRHT between irregularly shaped particles by developing a numerical model based on fluctuational electrodynamics [73], called the discrete system Green's function (DSGF) approach. The DSGF approach is a volume integral equation method that can be applied to arbitrary three-dimensional geometries and differs from previous volume integral approaches, such as the thermal discrete dipole approximation (TDDA) [66,67,69], by defining all system interactions generally and independent of the physics distinct to thermal excitation. The DSGF method is similar to the many-body method of NFRHT [55,56] in that interactions are defined by a generalized system Green's function. However, the DSGF approach is derived for and applicable to discretized objects, whereas the many-body method is limited to collections of particles modeled as independent dipoles. The DSGF method is an improvement over the other volume integral approach, the TDDA, because the system Green's function found in the DSGF method can be post-processed to calculate many quantities of interest to NFRHT, such as power dissipation and the local density of states. In contrast, the TDDA is limited to solving for the autocorrelation of total dipole moments, a more specialized parameter with less post-processing range. Additionally, the main matrix equation in the DSGF method is amenable to solution by a wider variety of computational algorithms than the main matrix equation in the TDDA.

In this paper, we apply the DSGF method to irregularly shaped dielectric particles made of $SiO_2$ and morphology characterized by Gaussian random spheres. We find that geometric irregularity significantly affects the spectrally integrated conductance for closely spaced particles with size approximately equal to or smaller than the vacuum separation distance, whereas resonance damping and broadening arise for all modeled separation distances. Since resonance broadening and damping appear for all the tested separation distances, this work is of particular interest for the design of thermal metamaterials composed of complex-shaped, particle-like substructures or to systems that deal with real-world particles containing defects.

The rest of the paper is organized as follows. First, we introduce the DSGF method for predicting NFRHT between three-dimensional objects of arbitrary shape (Sec. II). Next, we verify the DSGF method against the analytical solution for chains of two and three spheres (Sec. III). After verification, we apply the DSGF approach to model NFRHT between irregularly shaped dielectric particles with morphology defined by Gaussian random spheres (Sec. IV). Concluding remarks are presented in Sec. V.

## II. DESCRIPTION OF THE DISCRETE SYSTEM GREEN'S FUNCTION (DSGF) FORMALISM

The DSGF method is based on fluctuational electrodynamics [73,74] and is defined for a system of three-dimensional, thermally emitting objects of arbitrary number, geometry, size, and material embedded within a lossless background reference medium (FIG 1). The objects, occupying a total combined volume $V_{therm}$, are assumed to be in local thermodynamic equilibrium, lossy, nonmagnetic, and may have nonuniform dielectric function $\varepsilon(\mathbf{r}, \omega)$ and nonuniform temperature $T(\mathbf{r})$. Here, $\mathbf{r}$ is the position coordinate and $\omega$ is the angular frequency. The lossless background reference medium encompassing volume $V_{ref}$ may be vacuum or any other material for which the dielectric function $\varepsilon_{ref}(\omega)$ is strictly real-valued. The dielectric functions of both the background reference medium and the thermal objects are assumed to be isotropic and linear; however, the DSGF method may also be generalized to anisotropic materials with tensor dielectric functions.

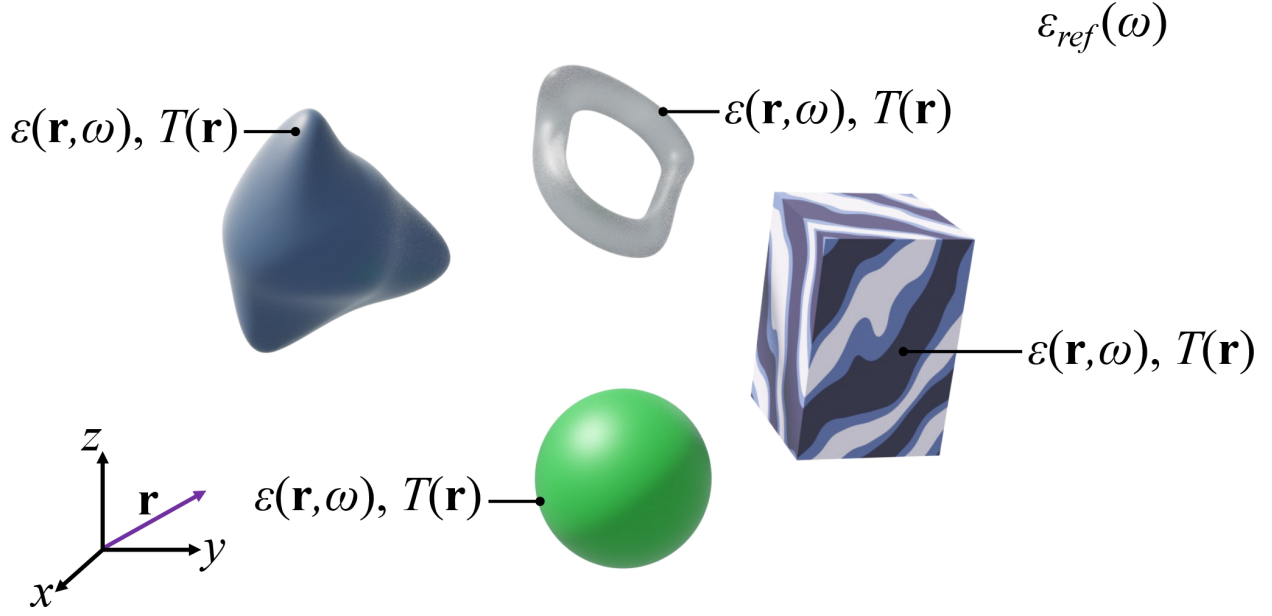

FIG 1. System of thermal objects of arbitrary number, geometry, size, and material occupying a total volume $V_{therm}$. The thermal objects may be of nonuniform temperature $T(\mathbf{r})$ and nonuniform dielectric function $\varepsilon(\mathbf{r},\omega)$. The thermal objects are embedded in a lossless background reference medium of volume $V_{ref}$ that is characterized by a real-valued dielectric function $\varepsilon_{ref}(\omega)$.

## A. Definition of the system Green's function

The DSGF method is derived from the stochastic Maxwell equations of fluctuational electrodynamics [73]. Separate wave equations are constructed for the thermal object domain and the background reference medium:

$$\nabla \times \nabla \times \mathbf{E}(\mathbf{r},\omega) - k_0^2 \varepsilon_{ref}(\omega)\mathbf{E}(\mathbf{r},\omega) = \mathbf{0}, \quad \mathbf{r} \in V_{ref}, \tag{1}$$

$$\nabla \times \nabla \times \mathbf{E}(\mathbf{r},\omega) - k_0^2 \varepsilon(\mathbf{r},\omega)\mathbf{E}(\mathbf{r},\omega) = i\omega\mu_0 \mathbf{J}^{(fl)}(\mathbf{r},\omega), \quad \mathbf{r} \in V_{therm}, \tag{2}$$

where $\mathbf{E}$ is the total electric field, $\mathbf{J}^{(fl)}$ is the fluctuating thermal source current density, $\mu_0$ is the vacuum permeability, and $k_0$ is the vacuum wavevector magnitude defined as $k_0 = \omega\sqrt{\mu_0 \varepsilon_0}$ with $\varepsilon_0$ being the vacuum permittivity. In Eqs. (1) and (2), the fields are assumed to be time-harmonic and vary as $e^{-i\omega t}$. Combining Eqs. (1) and (2), a wave equation that is valid over all real space $\mathfrak{R}^3$ is derived,

$$\nabla \times \nabla \times \mathbf{E}(\mathbf{r},\omega) - k_0^2\big[\varepsilon_{ref}(\omega) + \varepsilon_r(\mathbf{r},\omega)\big]\mathbf{E}(\mathbf{r},\omega) = i\omega\mu_0 \mathbf{J}^{(eq)}(\mathbf{r},\omega), \quad \mathbf{r} \in \mathfrak{R}^3, \tag{3}$$

with an equivalent current density $\mathbf{J}^{(eq)}$ defined as

$$\mathbf{J}^{(eq)}(\mathbf{r},\omega) = \begin{cases} \mathbf{0}, & \mathbf{r} \in V_{ref} \\ \mathbf{J}^{(fl)}(\mathbf{r},\omega), & \mathbf{r} \in V_{therm} \end{cases}, \tag{4}$$

where $\varepsilon_r(\mathbf{r},\omega)$ is the relative dielectric function with respect to the background reference medium,

$$\varepsilon_r(\mathbf{r},\omega) = \begin{cases} 0, & \mathbf{r} \in V_{ref} \\ \varepsilon(\mathbf{r},\omega) - \varepsilon_{ref}(\omega), & \mathbf{r} \in V_{therm} \end{cases}. \tag{5}$$

The electric field solution to the wave equation (3) is determined using the volume integral technique. The total electric field is the sum of the homogeneous and particular solutions

$$\mathbf{E}(\mathbf{r},\omega) = \mathbf{E}_0(\mathbf{r},\omega) + i\omega\mu_0 \int_V \bar{\bar{\mathbf{G}}}(\mathbf{r},\mathbf{r}',\omega)\, \mathbf{J}^{(eq)}(\mathbf{r}',\omega) d^3\mathbf{r}', \quad \mathbf{r} \in \mathfrak{R}^3, \tag{6}$$

where $\mathbf{E}_0(\mathbf{r},\omega)$ is the homogeneous solution, the integral term is the particular solution, and $\bar{\bar{\mathbf{G}}}(\mathbf{r},\mathbf{r}',\omega)$ is the system Green's function, also sometimes called the dyadic Green's function. The homogeneous solution $\mathbf{E}_0(\mathbf{r},\omega)$ is the field that exists due to the objects and reference medium when fluctuating thermal sources are not present and satisfies the wave equation

$$\nabla \times \nabla \times \mathbf{E}_0(\mathbf{r},\omega) - k_0^2\big[\varepsilon_{ref}(\omega) + \varepsilon_r(\mathbf{r},\omega)\big]\mathbf{E}_0(\mathbf{r},\omega) = \mathbf{0}, \quad \mathbf{r} \in \mathfrak{R}^3. \tag{7}$$

Specifically, Eq. (7) describes electromagnetic scattering by imposed external fields, such as laser excitation or irradiation by the surroundings, rather than imposed source current densities. In this work, we focus on the case where there are no imposed fields, such that $\mathbf{E}_0(\mathbf{r},\omega) = \mathbf{0}$. For application of the DSGF method to systems with external field excitation, the full solution to Eq. (7) is provided in Appendix A. In the absence of imposed fields, Eq. (6) simplifies as

$$\mathbf{E}(\mathbf{r},\omega) = i\omega\mu_0 \int_{V_{therm}} \bar{\bar{\mathbf{G}}}(\mathbf{r},\mathbf{r}',\omega)\mathbf{J}^{(fl)}(\mathbf{r}',\omega)\, d^3\mathbf{r}', \quad \mathbf{r} \in \mathfrak{R}^3, \tag{8}$$

where the condition that $\mathbf{J}^{(eq)}(\mathbf{r}', \omega) = \mathbf{0}$ for $\mathbf{r}' \in V_{ref}$ has been used to restrict the integral to the thermal object domain. The system Green's function $\bar{\bar{\mathbf{G}}}(\mathbf{r}, \mathbf{r}', \omega)$, relating the electric field observed at point $\mathbf{r}$ due to a point source excitation at location $\mathbf{r}'$ for a given frequency $\omega$, completely and deterministically defines the electromagnetic response of the system to any imposed sources. Specifically, the system Green's function satisfies the original wave equation (Eq. (3)) for a point excitation, $\bar{\bar{\mathbf{I}}}\delta(\mathbf{r} - \mathbf{r}')$

$$\nabla \times \nabla \times \bar{\bar{\mathbf{G}}}(\mathbf{r}, \mathbf{r}', \omega) - k_0^2 [\varepsilon_{ref}(\omega) + \varepsilon_r(\mathbf{r}, \omega)] \bar{\bar{\mathbf{G}}}(\mathbf{r}, \mathbf{r}', \omega) = \bar{\bar{\mathbf{I}}}\delta(\mathbf{r} - \mathbf{r}'), \quad \mathbf{r} \in \Re^3, \tag{9}$$

where $\bar{\bar{\mathbf{I}}}$ is the unit dyadic and $\delta$ is the Dirac function.

Eq. (8) is the typical starting point for analytical derivations of NFRHT where closed-form expressions of the system Green's function may be attained through implementation of appropriate boundary conditions. Due to prohibitive mathematical complexity, analytical solutions have been restricted to simple geometries, such as layered media [75] and spheres [52–54]. In the DSGF method, the system Green's function for arbitrarily shaped three-dimensional objects is calculated numerically.

**B. Derivation of self-consistent system Green's function equation**

The system Green's function can be expressed in terms of the free-space Green's function $\bar{\bar{\mathbf{G}}}^0(\mathbf{r}, \mathbf{r}', \omega)$, which has known analytical solution [76,77]. The free-space Green's function describes the response to excitation of an infinite, lossless, homogeneous medium, such as the background reference medium presented here, and satisfies the wave equation

$$\nabla \times \nabla \times \bar{\bar{\mathbf{G}}}^0(\mathbf{r}, \mathbf{r}', \omega) - k_0^2 \varepsilon_{ref}(\omega) \bar{\bar{\mathbf{G}}}^0(\mathbf{r}, \mathbf{r}', \omega) = \bar{\bar{\mathbf{I}}}\delta(\mathbf{r} - \mathbf{r}'), \quad \mathbf{r} \in \Re^3. \tag{10}$$

To facilitate clarity of mathematical procedures, the free-space Green's function wave equation (Eq. (10)) and the system Green's function wave equation (Eq. (9)) are transformed from standard $\mathbf{r}$ representation into operator notation

$$\left(\mathbb{L} + \mathbb{e}_{ref}\right)\mathbb{G}^0 = \mathbb{I}, \tag{11}$$

$$\left(\mathbb{L} + \mathbb{e}_{ref} + \mathbb{e}_r\right)\mathbb{G} = \mathbb{I}, \tag{12}$$

where operators are defined as $\mathbb{L} \triangleq \nabla \times \nabla \times$, $\mathbb{e}_{ref} \triangleq -k_0^2 \varepsilon_{ref}(\omega)$, $\mathbb{e}_r \triangleq -k_0^2 \varepsilon_r(\mathbf{r}, \omega)$, $\mathbb{G} \triangleq \bar{\bar{\mathbf{G}}}(\mathbf{r}, \mathbf{r}', \omega)$, $\mathbb{G}^0 \triangleq \bar{\bar{\mathbf{G}}}^0(\mathbf{r}, \mathbf{r}', \omega)$, and $\mathbb{I} \triangleq \bar{\bar{\mathbf{I}}}\delta(\mathbf{r} - \mathbf{r}')$. This is similar to the method used in Ref. [78]. Combining Eqs. (11) and (12), and recognizing that Eq. (11) can be rearranged as $\left(\mathbb{L} + \mathbb{e}_{ref}\right) = (\mathbb{G}^0)^{-1}$, the system Green's function can be described as

$$\mathbb{G} = \mathbb{G}^0 - \mathbb{G}^0 \mathbb{e}_r \mathbb{G}. \tag{13}$$

This is a self-consistent equation for the system Green's function, analogous to the Dyson equation of quantum field theory [77–80]. After slightly rearranging Eq. (13) and converting back to standard **r** representation [81,82], we obtain

$$\bar{\bar{\mathbf{G}}}^0(\mathbf{r}, \mathbf{r}', \omega) = \bar{\bar{\mathbf{G}}}(\mathbf{r}, \mathbf{r}', \omega) - k_0^2 \int_{V_{therm}} \bar{\bar{\mathbf{G}}}^0(\mathbf{r}, \mathbf{r}'', \omega) \varepsilon_r(\mathbf{r}'', \omega) \bar{\bar{\mathbf{G}}}(\mathbf{r}'', \mathbf{r}', \omega) d^3 \mathbf{r}''. \tag{14}$$

In this formulation, the response of the system is defined independently of the physics of the imposed source, and the parameters in Eq. (14) are all deterministic.

### C. Discretization

To solve the self-consistent system Green's function equation (Eq. (14)), the volume of the thermal objects $V_{therm}$ is discretized into a total of $N$ cubic subvolumes along a cubic lattice, where $\mathbf{r}_i$ is the center point of each discretized subvolume $\Delta V_i$ and $i = 1, 2, \ldots, N$. Each subvolume is assumed to have uniform dielectric function, electric field, Green's functions, and temperature. Determining the appropriate subvolume size for a given system is discussed in detail in Ref. [67]. Two general rules for proper discretization are that subvolume size should be smaller than the inverse magnitude of the wavevector in the thermal objects, $(\Delta V)^{1/3} \ll 1/k$, where $k =$ $\omega\sqrt{\varepsilon(\mathbf{r},\omega)\varepsilon_0\mu_0}$, and the size of the subvolumes must be small compared to the vacuum separation distance between the thermal objects, $(\Delta V)^{1/3} \ll d$.

Following the general procedure described in Ref. [67], Eq. (14) is discretized as

$$\int_{\Delta V_i}\int_{\Delta V_j} \bar{\bar{\mathbf{G}}}^0(\mathbf{r},\mathbf{r}',\omega)\, d^3\mathbf{r}'d^3\mathbf{r}$$
$$= \int_{\Delta V_i}\int_{\Delta V_j} \bar{\bar{\mathbf{G}}}(\mathbf{r},\mathbf{r}',\omega)\, d^3\mathbf{r}'d^3\mathbf{r}$$
$$- k_0^2 \int_{\Delta V_i}\int_{\Delta V_j}\int_{V_{therm}} \bar{\bar{\mathbf{G}}}^0(\mathbf{r},\mathbf{r}'',\omega)\varepsilon_r(\mathbf{r}'',\omega)\bar{\bar{\mathbf{G}}}(\mathbf{r}'',\mathbf{r}',\omega)\, d^3\mathbf{r}''\, d^3\mathbf{r}'d^3\mathbf{r}.$$

(15)

Assuming that $\bar{\bar{\mathbf{G}}}$ is well-behaved over the entire domain and applying the principal value approximation for the singularity in $\bar{\bar{\mathbf{G}}}^0$ at $\mathbf{r} = \mathbf{r}'$ [83,84], Eq. (15) is simplified as

$$\bar{\bar{\mathbf{G}}}^0(\mathbf{r}_i,\mathbf{r}_j,\omega) = \bar{\bar{\mathbf{G}}}(\mathbf{r}_i,\mathbf{r}_j,\omega) - k_0^2 \sum_{k=1}^{N} \bar{\bar{\mathbf{G}}}^0(\mathbf{r}_i,\mathbf{r}_k,\omega)\Delta V_k \varepsilon_r(\mathbf{r}_k,\omega)\bar{\bar{\mathbf{G}}}(\mathbf{r}_k,\mathbf{r}_j,\omega). \qquad (16)$$

The discretized version of the known free-space Green's function $\bar{\bar{\mathbf{G}}}^0(\mathbf{r}_i,\mathbf{r}_j,\omega)$ is presented in Appendix B. The full system of equations describing all interactions between subvolumes may be expanded in matrix form as

$$\left\{ \begin{bmatrix} \bar{\bar{\mathbf{I}}} & 0 & 0 \\ 0 & \ddots & 0 \\ 0 & 0 & \bar{\bar{\mathbf{I}}} \end{bmatrix} - k_0^2 \begin{bmatrix} \bar{\bar{\mathbf{G}}}^0_{11} & \cdots & \bar{\bar{\mathbf{G}}}^0_{1N} \\ \vdots & \ddots & \vdots \\ \bar{\bar{\mathbf{G}}}^0_{N1} & \cdots & \bar{\bar{\mathbf{G}}}^0_{NN} \end{bmatrix} \begin{bmatrix} \alpha_1^{(0)} & 0 & 0 \\ 0 & \ddots & 0 \\ 0 & 0 & \alpha_N^{(0)} \end{bmatrix} \right\} \begin{bmatrix} \bar{\bar{\mathbf{G}}}_{11} & \cdots & \bar{\bar{\mathbf{G}}}_{1N} \\ \vdots & \ddots & \vdots \\ \bar{\bar{\mathbf{G}}}_{N1} & \cdots & \bar{\bar{\mathbf{G}}}_{NN} \end{bmatrix}$$
$$= \begin{bmatrix} \bar{\bar{\mathbf{G}}}^0_{11} & \cdots & \bar{\bar{\mathbf{G}}}^0_{1N} \\ \vdots & \ddots & \vdots \\ \bar{\bar{\mathbf{G}}}^0_{N1} & \cdots & \bar{\bar{\mathbf{G}}}^0_{NN} \end{bmatrix}, \qquad (17)$$

where lattice locations $\mathbf{r}_i$ and $\mathbf{r}_j$ are represented by subscripts $i$ and $j$, and the bare polarizability is defined as $\alpha_k^{(0)} = \Delta V_k \varepsilon_r(\mathbf{r}_k,\omega)$. Eq. (17) is a system of linear equations of the form $\bar{\bar{\mathbf{A}}}\bar{\bar{\mathbf{G}}} = \bar{\bar{\mathbf{G}}}^0$, where $\bar{\bar{\mathbf{G}}}$ represents the DSGF and $\bar{\bar{\mathbf{A}}}$ is the interaction matrix defined by the term in curly brackets.

The full matrices, $\bar{\bar{\mathbf{A}}}$, $\bar{\bar{\mathbf{G}}}$, and $\bar{\bar{\mathbf{G}}}^0$ each contain a total of $3N \times 3N$ terms, accounting for $N$ subvolumes and three Cartesian vector components. Each submatrix, $\bar{\bar{\mathbf{A}}}_{ij}$, $\bar{\bar{\mathbf{G}}}^0_{ij}$, and $\bar{\bar{\mathbf{G}}}_{ij}$, is a $3 \times 3$ matrix that describes the interaction between the $i$th and $j$th subvolumes. The many-body approach for NFRHT presented in Refs. [55,56] defines a similar self-consistent Green's function equation. However, the many-body approach assumes that particles are within the dipole regime. Here, the DSGF formulation is general and can be applied to thermal objects of any size and shape, without restriction to their separation distance, as long as proper discretization is heeded.

The advantage of the DSGF method over other volume integral methods, such as the TDDA, is that the matrix equation presented in Eq. (17) is of the well-known form $\bar{\bar{\mathbf{A}}}\bar{\bar{\mathbf{X}}} = \bar{\bar{\mathbf{B}}}$ for which there are a variety of solution algorithms and open-source solvers available. The DSGF matrix equation defined in Eq. (17) can be solved directly using LU decomposition to invert $\bar{\bar{\mathbf{A}}}$, or it can be solved with iterative techniques, such as the biconjugate gradient stabilized method, the conjugate gradient squared method, or the perturbative-like technique presented in Ref. [78]. Iterative techniques have the advantage of reducing computational memory requirements, which can become prohibitive for large numbers of subvolumes [67]. Conversely, the main matrix equation of the TDDA is of the form $\bar{\bar{\mathbf{A}}}\bar{\bar{\mathbf{X}}}\bar{\bar{\mathbf{A}}}^\dagger = \bar{\bar{\mathbf{B}}}$ and has fewer documented solution methods. More details on the comparison of the matrix equations solved in the DSGF and TDDA methods can be found in Appendix C. Another advantage of the DSGF method over the TDDA is that the variable of interest in the DSGF method is the system Green's function, a general parameter that is independent of the physics of thermal excitation and that can be post-processed to solve for a variety of other thermal parameters. The variable of interest $\bar{\bar{\mathbf{X}}}$ in the TDDA method is the autocorrelation of the total dipole moment, a more restrictive parameter that combines information of thermal excitation and scattering.

## D. Radiative heat transfer

Once the system Green's function is known via the solution of the system of equations (17), the net radiative heat transfer can be calculated. Using Poynting's theorem, the net power dissipated in the thermal objects is defined as [73,85]

$$\langle Q_{V_{therm}}(t)\rangle = \int_{V_{therm}} \langle \boldsymbol{\mathcal{E}}(\mathbf{r},t) \cdot \left(\boldsymbol{\mathcal{J}}^{(fl)}(\mathbf{r},t) + [\tilde{\sigma} * \boldsymbol{\mathcal{E}}(\mathbf{r},t)]\right)\rangle d^3\mathbf{r}, \tag{18}$$

where $t$ is the time offset, $\tilde{\sigma}$ is the instantaneous material conductivity, $\boldsymbol{\mathcal{E}}(\mathbf{r},t)$ is the time-dependent instantaneous electric field, $\boldsymbol{\mathcal{J}}^{(fl)}(\mathbf{r},t)$ is the time-dependent fluctuating thermal current density, the $*$ symbol represents the convolution operation, the $\cdot$ symbol represents the dot product, and $\langle \ \rangle$ brackets represent the ensemble average. The fluctuating thermal current density $\boldsymbol{\mathcal{J}}^{(fl)}(\mathbf{r},t)$ is defined as a wide-sense stationary random process and embodies the statistical nature of thermal excitation of charged microparticles [73,74]. For such processes, the first moment (i.e., mean) of the stochastic signal is zero, $\langle \boldsymbol{\mathcal{J}}^{(fl)}(\mathbf{r},\tau)\rangle = \mathbf{0}$, and the second moment (i.e., autocorrelation) is nonzero, $\langle \boldsymbol{\mathcal{J}}^{(fl)}(\mathbf{r},\tau)\boldsymbol{\mathcal{J}}^{(fl)\dagger}(\mathbf{r},\tau-t)\rangle \neq \mathbf{0}$ [86]. The autocorrelation function of the fluctuating thermal current density is defined by the fluctuation-dissipation theorem, which relates equilibrium fluctuations in current density to macroscale electromagnetic dissipation inside a thermal object. In its spectral representation, the fluctuation-dissipation theorem is expressed as [73,87]

$$\langle \mathbf{J}^{(fl)}(\mathbf{r},\omega)\mathbf{J}^{(fl)\dagger}(\mathbf{r}',\omega')\rangle = 4\pi\omega\varepsilon_0 \text{Im}[\varepsilon(\mathbf{r},\omega)]\Theta(\omega,T)\delta(\mathbf{r}-\mathbf{r}')\delta(\omega-\omega')\bar{\bar{\mathbf{I}}}, \tag{19}$$

where the mean energy of an electromagnetic state is defined as $\Theta(\omega,T) = \hbar\omega\left(e^{\frac{\hbar\omega}{k_B T}} - 1\right)^{-1}$, $\hbar$ is the reduced Planck constant, and $k_B$ is the Boltzmann constant.

Using Eq. (8) to describe the electric field in terms of the system Green's function and incorporating the fluctuation-dissipation theorem, the net power dissipated given by Eq. (18) becomes

$$\langle Q_{V_{therm}}(t)\rangle = \int_{V_{therm}} \frac{2}{\pi} \int_0^\infty k_0^2 \left\{ \mathrm{Im}[\varepsilon(\mathbf{r},\omega)]\Theta(\omega,T)\mathrm{Im}\left(\mathrm{Tr}[\bar{\bar{\mathbf{G}}}^\dagger(\mathbf{r},\mathbf{r},\omega)]\right) + \right.$$

$$\left. k_0^2 \mathrm{Im}[\varepsilon(\mathbf{r},\omega)] \int_{V_{therm}} \mathrm{Im}[\varepsilon(\mathbf{r}',\omega)]\Theta(\omega,T')\mathrm{Tr}[\bar{\bar{\mathbf{G}}}(\mathbf{r},\mathbf{r}',\omega)\bar{\bar{\mathbf{G}}}^\dagger(\mathbf{r},\mathbf{r}',\omega)] d^3\mathbf{r}' \right\} d\omega d^3\mathbf{r}, \qquad (20)$$

where Tr represents the trace. In Eq. (20), the inverse Fourier transform convention $f(t) = \frac{1}{2\pi}\int_{-\infty}^{+\infty} F(\omega)e^{-i\omega t}d\omega$ is followed. Discretizing Eq. (20) along a cubic lattice, the net power dissipated in a subvolume $\Delta V_i$ is defined as

$$\langle Q_{\Delta V_i}(t)\rangle = \frac{2}{\pi}\int_0^\infty k_0^4 \Delta V_i \mathrm{Im}[\varepsilon(\mathbf{r}_i,\omega)] \left\{ \sum_{\substack{j=1 \\ j\neq i}}^{N} \Delta V_j \mathrm{Im}[\varepsilon(\mathbf{r}_j,\omega)][\Theta(\omega,T_j) - \right.$$

$$\left. \Theta(\omega,T_i)]\mathrm{Tr}[\bar{\bar{\mathbf{G}}}(\mathbf{r}_i,\mathbf{r}_j,\omega)\bar{\bar{\mathbf{G}}}^\dagger(\mathbf{r}_i,\mathbf{r}_j,\omega)] \right\} d\omega. \qquad (21)$$

In deriving Eq. (21), the condition of zero power dissipation at thermal equilibrium is implemented such that $\langle Q_{\Delta V_i}(t)\rangle = 0$ when $\Theta(\omega,T_i) = \Theta(\omega,T_j)$.

Rearranging into a Landauer-like form, the net power dissipated in a subvolume is simplified as

$$\langle Q_{\Delta V_i}(t)\rangle = \frac{1}{2\pi}\int_0^\infty \left\{ \sum_{\substack{j=1 \\ j\neq i}}^{N} [\Theta(\omega,T_j) - \Theta(\omega,T_i)]\mathcal{T}_{ij}(\omega) \right\} d\omega, \qquad (22)$$

where the transmission coefficient between subvolumes $i$ and $j$ is defined as

$$\mathcal{T}_{ij}(\omega) = 4k_0^4 \Delta V_i \Delta V_j \mathrm{Im}[\varepsilon(\mathbf{r}_i,\omega)]\mathrm{Im}[\varepsilon(\mathbf{r}_j,\omega)]\mathrm{Tr}[\bar{\bar{\mathbf{G}}}(\mathbf{r}_i,\mathbf{r}_j,\omega)\bar{\bar{\mathbf{G}}}^\dagger(\mathbf{r}_i,\mathbf{r}_j,\omega)]. \qquad (23)$$

The net power dissipated in a bulk thermal object occupying the closed volume $V_A$ is given by

$$\langle Q_A(t)\rangle = \frac{1}{2\pi}\int_0^\infty \sum_{i\in V_A}\left\{ \sum_{j\notin V_A}[\Theta(\omega,T_j) - \Theta(\omega,T_i)]\mathcal{T}_{ij}(\omega) \right\} d\omega. \qquad (24)$$

From Eq. (23), it can be seen that the transmission coefficient between two subvolumes is dependent on the trace of the product of the DSGF and its conjugate transpose (i.e., the squared Frobenius norm of the DSGF). In this framework, solving a NFRHT problem reduces to solving for the appropriate system Green's function, from which the net heat rate can be easily calculated.

Another quantity used in characterizing NFRHT is the conductance. The spectral conductance is defined as the power transfer between two objects for a given frequency in the limit that their temperature difference $\delta T$ goes to zero. Mathematically, this is expressed for bulk objects $A$ and $B$ as [54]

$$G_{AB}(\omega, T) = \lim_{\delta T \to 0} \frac{\langle Q_{AB}(\omega) \rangle}{\delta T} = \left[\frac{\partial \Theta(\omega, T')}{\partial T}\right]_{T'=T} \mathcal{T}_{AB}(\omega), \qquad (25)$$

where the spectral transmission coefficient between the two bulk objects enclosing volumes $V_A$ and $V_B$ is calculated as

$$\mathcal{T}_{AB}(\omega) = \sum_{i \in V_A} \sum_{j \in V_B} \mathcal{T}_{ij}(\omega). \qquad (26)$$

Reciprocity holds such that $\mathcal{T}_{AB}(\omega) = \mathcal{T}_{BA}(\omega)$. The total conductance at a given temperature $T$ between two bulk objects is then determined by integration over all frequencies, $G_{t,AB}(T) = \frac{1}{2\pi} \int_0^\infty G_{AB}(\omega, T) \, d\omega$. The spectral and total conductance are used to characterize NFRHT between dielectric particles in the following sections.

### III. VERIFICATION OF THE DISCRETE SYSTEM GREEN'S FUNCTION (DSGF) METHOD AGAINST THE ANALYTICAL SOLUTION FOR SPHERES

To determine the accuracy of the DSGF method in modeling NFRHT between particles, we compare DSGF calculations of conductance at room temperature ($T = 300$ K) against the analytical solution for chains of two and three dielectric spheres made of $SiO_2$ embedded in vacuum. The analytical solution is calculated using the technique from Ref. [54]. The dielectric function of $SiO_2$, taken from Ref. [54], is provided in Sec. S1 of the Supplemental Material [88].

The discretized spheres used in the DSGF method are modeled to have the same volume and center-of-mass separation distance as the corresponding perfect spheres of the analytical solution. The accuracy of the DSGF method is evaluated for vacuum separation distances $d$ ranging from 10 nm to 10 μm and for two different sphere radii, $R = 50$ nm and $R = 500$ nm (FIG 2).

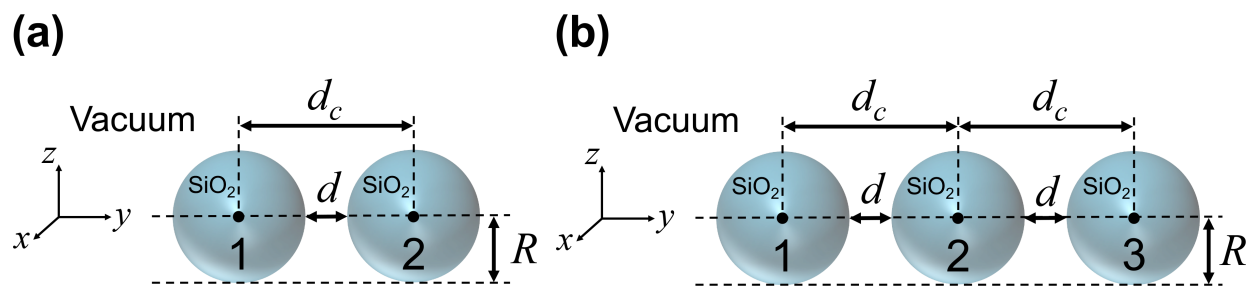

FIG 2. Schematics of (a) two SiO$_2$ spheres and (b) a linear chain of three SiO$_2$ spheres embedded in vacuum. Two different sphere sizes are considered: $R = 50$ nm and $R = 500$ nm. The vacuum separation distance varies from 10 nm $\leq d \leq$ 10 μm. The symbol $d_c$ represents the center-of-mass separation distance.

### A. Two spheres

The total conductance calculated by the DSGF method shows good agreement with the analytical solution for two spheres (FIG 3). For vacuum separation distances in the range $R \lesssim d \lesssim 20R$, the absolute value of the error between the DSGF method and analytical solution is below 3% for 50-nm-radius spheres and below 5% for 500-nm-radius spheres. The slight increase in error at the closest and farthest vacuum separation distances was expected from the TDDA literature [67], since the DSGF method and TDDA have equivalent convergence behavior and generate equivalent values of conductance when the same discretization is used (see Sec. S2 of the Supplemental Material [88]).

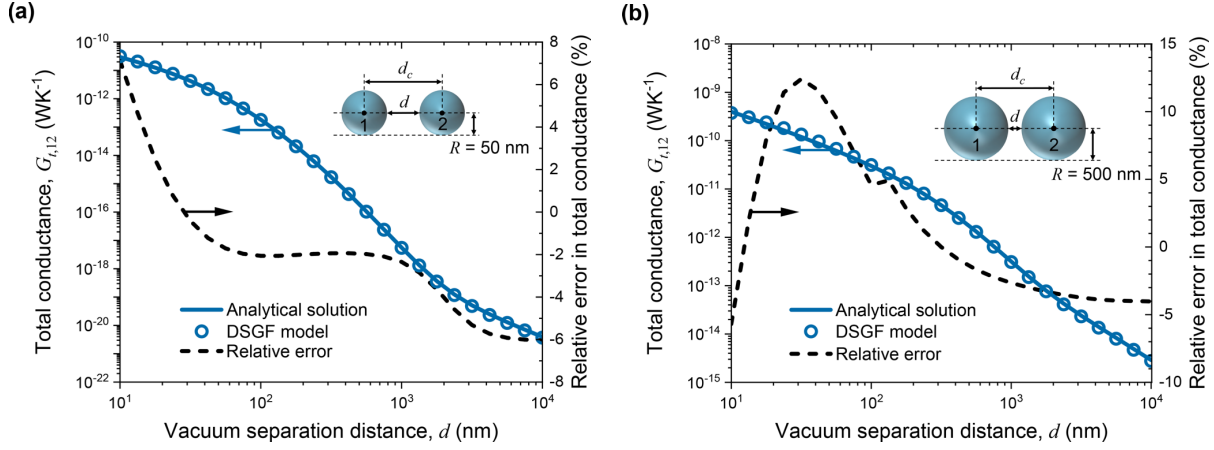

FIG 3. Comparison of the total conductance at $T = 300$ K between two SiO$_2$ spheres calculated analytically [54] and by the DSGF method. Spheres of radii (a) $R = 50$ nm and (b) $R = 500$ nm are modeled for variable vacuum separation distance, 10 nm $\leq d \leq$ 10 μm. 50-nm-radius spheres are discretized into $N_{sphere} = 2176$ subvolumes per sphere for all vacuum separation distances. 500-nm-radius spheres are discretized into $N_{sphere} = 5616$ subvolumes per sphere for $d \leq 100$ nm and into $N_{sphere} = 2176$ subvolumes per sphere for $d > 100$ nm. The relative error between the DSGF and analytical solution is calculated as $\frac{G_{t,12}^{DSGF} - G_{t,12}^{analytical}}{G_{t,12}^{analytical}}$.

For the 50-nm-radius sphere case (FIG 3(a)), increased error at small vacuum separation distances $d < R$ can be attributed to shape error (i.e., error due to approximating curved surfaces by cubic subvolumes). For the 500-nm-radius sphere case at small vacuum separation distances $d < R$ (FIG 3(b)), there is error due to approximating fields as uniform within a subvolume in addition to the aforementioned shape error. Shape error dominates for $R/10 \lesssim d < R$ (50 nm $\lesssim d <$ 500 nm for $R = 500$ nm), and nonuniform field error dominates cases for $d \lesssim R/10$ ($d \lesssim$ 50 nm for $R = 500$ nm). As the vacuum separation distance is reduced below the length scale of the subvolumes ($d < L_{sub}$, where $L_{sub} = (\Delta V)^{1/3}$), the fields within subvolumes at adjacent surfaces of the spheres display nonnegligible variation and can no longer be approximated as uniform. Since NFRHT between SiO$_2$ surfaces is dominated by SPhPs with penetration depth approximately equal to the vacuum separation distance $d$ [89], accurate description of the exponentially decaying field within the sphere requires that $d < L_{sub}$ [67]. The subvolume size for the 500-nm-radius spheres is $L_{sub} = 45.3$ nm for cases in which $d \leq 100$ nm, so the error trend changes when the vacuum separation distance is reduced below 45.3 nm. This can be seen in FIG 3(b) from the change in the error curve at the local maximum around a $d$ value of 42 nm. Note that for 50-nm-radius spheres, the vacuum separation distance never falls below the subvolume size ($L_{sub} = 6.2$ nm for all $d$ values), so this type of error does not arise in FIG 3(a). The error in the DSGF-calculated total conductance is still relatively low at small vacuum separation distances $d < R$ (less than 8% and 13% for $R$ = 50 nm and 500 nm, respectively). If increased accuracy is desired, both shape error and the error due to nonuniform fields within a subvolume can be reduced by refining the discretization with smaller subvolumes. Alternatively, when $d \ll R$, the proximity approximation [90], in which NFRHT is approximated as a summation of local conductance between two parallel surfaces with varying vacuum separation distances, is applicable provided that the spheres are optically thick [67]. When these conditions are satisfied, it may be simpler to calculate NFRHT between spheres via the proximity approximation than the DSGF method with refined discretization.

In FIG 3(a), the increase in error at large vacuum separation distances $d > 20R$ ($d > 1000$ nm) is due to additive numerical error stemming from over-discretization of the thermal objects [67]. This error may be reduced to $< 1\%$ by changing the discretization from many subvolumes to one subvolume per sphere (FIG 4). The DSGF method with one subvolume per object is equivalent to the many-body approach developed by Ben-Abdallah *et al.* [55,56] discussed in Sec. II. The many-body approach is similar to analytical dipole approximations of NFRHT [91] except that multiple scattering effects are included. Both the many-body approximation and the DSGF method with one subvolume per object are appropriate when the vacuum separation distance between objects is sufficiently large compared to the size of the objects and when the object size is much smaller than the thermal wavelength. For the 50-nm-radius spheres presented here, appropriate vacuum separation distances for application of the many-body approximation are $d > 20R$ ($d > 1000$ nm) (FIG 4).

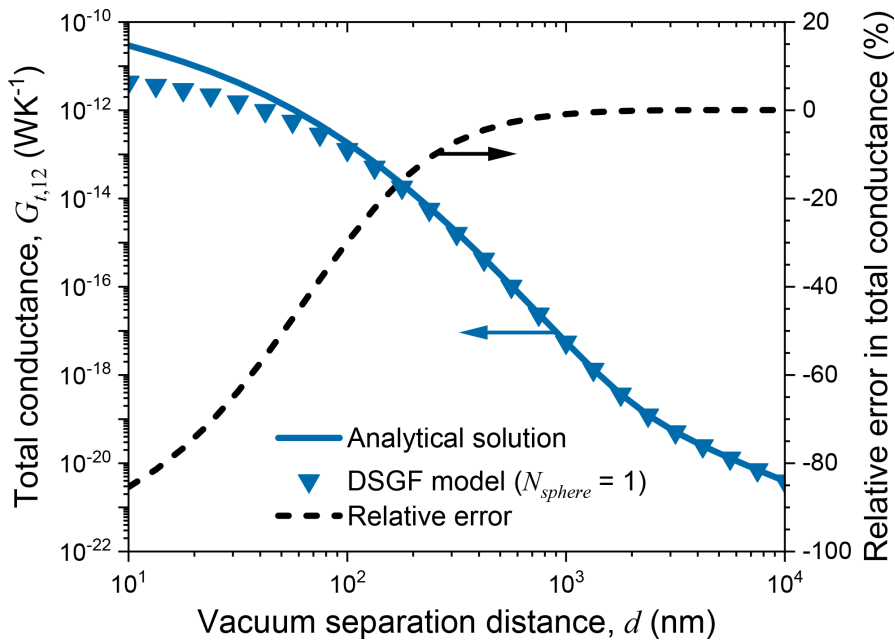

FIG 4. Comparison of the total conductance at $T = 300$ K between two 50-nm-radius SiO$_2$ spheres calculated analytically [54] and by the DSGF method when one subvolume per sphere is used in the discretization. Vacuum separation distance is varied as 10 nm $\leq d \leq$ 10 μm. The relative error between the DSGF and analytical solution is calculated as $\frac{G_{t,12}^{DSGF} - G_{t,12}^{analytical}}{G_{t,12}^{analytical}}$.

In addition to the total conductance, we verified that the DSGF method accurately captures the spectral behavior of NFRHT. Plots of spectral conductance for spheres of radii $R = 50$ nm and 500 nm and vacuum separation distances $d \approx R$ and $d \approx 20R$ are provided in Sec. S3 of the Supplemental Material [88].

## B. Linear chain of three spheres

In order to verify that the DSGF method may be extended to systems which include more than two objects, we model a linear chain of three dielectric spheres made of $SiO_2$. The same values of radius, vacuum separation distance, and subvolume size are used as those in the two-sphere analysis. Similar to the two-sphere cases, the DSGF method shows good agreement with the analytical solution for three spheres when proper discretization is applied (FIG 5). The total conductance between neighboring spheres ($G_{t,12}$) follows error trends similar to the two-sphere cases at close vacuum separation distances when over-discretization error is negligible. For the two outer spheres, the error in the total conductance ($G_{t,13}$) displays less variation than that for the neighboring spheres. The three-sphere error trends deviate from the two-sphere cases only at large vacuum separation distances ($d > 1000$ nm). At large vacuum separation distances, the over-discretization error in the three-sphere systems is greater than that for two spheres. This is expected from the addition of the third discretized sphere and increase in the total number of subvolumes. As before, this error at large vacuum separation distances can easily be reduced to less than 1% for both $G_{t,12}$ and $G_{t,13}$ when spheres are discretized into one subvolume per sphere (see Sec. S4 of the Supplemental Material [88]). From these results, we conclude that the DSGF method can accurately model NFRHT between multiple three-dimensional objects.

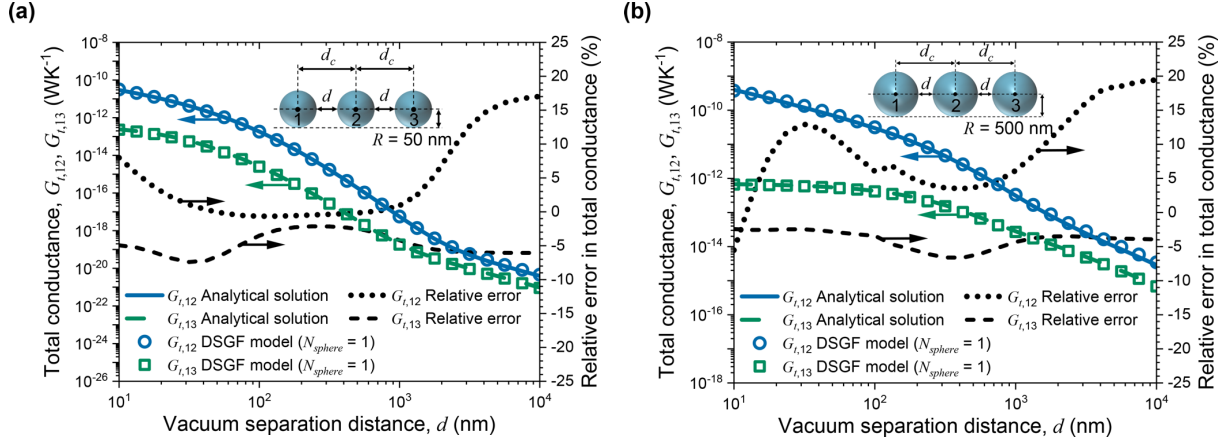

FIG 5. Comparison of the total conductance at $T = 300$ K between three SiO$_2$ spheres calculated analytically [54] and by the DSGF method. Linear chain of spheres of radii (a) $R = 50$ nm and (b) $R = 500$ nm are modeled for variable vacuum separation distance, $10 \text{ nm} \leq d \leq 10 \text{ μm}$. 50-nm-radius spheres are discretized into $N_{sphere} = 2176$ subvolumes per sphere for all vacuum separation distances. 500-nm-radius spheres are discretized into $N_{sphere} = 5616$ subvolumes per sphere for $d \leq 100$ nm and into $N_{sphere} = 2176$ subvolumes per sphere for $d > 100 \text{ nm}$. The relative error between the DSGF and analytical solution is calculated as $\frac{G_{t,12}^{DSGF} - G_{t,12}^{analytical}}{G_{t,12}^{analytical}}$ and $\frac{G_{t,13}^{DSGF} - G_{t,13}^{analytical}}{G_{t,13}^{analytical}}$.

From all the verification checks for two and three spheres presented in Sec. III, the DSGF method is deemed suitable for the study of NFRHT between irregularly shaped particles for which no analytical solutions exist. NFRHT between irregularly shaped dielectric particles is discussed next.

## IV. NEAR-FIELD RADIATIVE HEAT TRANSFER BETWEEN IRREGULARLY SHAPED DIELECTRIC PARTICLES

We apply the DSGF method to model NFRHT between two irregularly shaped SiO$_2$ particles embedded in vacuum. Particle dimensions, discretizations, and separation distances are chosen to minimize numerical error in the DSGF model. Based on verification results for 50-nm-radius spheres, minimal numerical error is expected for subvolume lengths $L_{sub} \lesssim 6$ nm, and vacuum separation distances in the range $4L_{sub} \lesssim d \lesssim 20R$. As such, all irregularly shaped particles are modeled with subvolume lengths $L_{sub} = 3.8$ nm and equivalent radii in the range 35 nm $< R_{eq} <$ 50 nm. Equivalent radii are defined as the radii of perfect spheres of equivalent volume as the irregularly shaped particles. Since the vacuum separation distance $d$ of irregularly shaped particles varies with particle orientation, we use the center-of-mass separation distance $d_c$ to describe particle systems in this section. The center-of-mass separation distance is independent of particle orientation and is therefore a more general parameter for calculating NFRHT trends. We compare the conductance calculated between spheres and irregularly shaped particles of equivalent center-of-mass separation distance $d_c$. The corresponding range of center-of-mass separation distances that ensures minimal error is $(4L_{sub} + R_{eq,1} + R_{eq,2}) \lesssim d_c \lesssim (20 \times \min[R_{eq,1}, R_{eq,2}] + R_{eq,1} + R_{eq,2})$.

We model two types of irregularly shaped particles: particles with mild distortion and particles with high distortion from perfect spherical geometry. Particle distortion is quantified using the Gaussian random sphere method [92–94]. In this approach, the particle surface is defined with a randomly weighted linear combination of spherical harmonics. The radius and real-valued logarithmic radius of the Gaussian random sphere are, respectively,

$$r(\theta, \varphi) = \frac{a \exp[s(\theta,\varphi)]}{\sqrt{1+\sigma^2}} \tag{27}$$

and

$$s(\theta, \varphi) = \sum_{\ell=0}^{\infty} \sum_{m=-\ell}^{\ell} s_{\ell m} Y_\ell^m(\theta, \varphi), \tag{28}$$

where $\theta$ is the polar angle, $\varphi$ is the azimuthal angle, $a$ is the mean radius, $\sigma$ is the relative standard deviation of the radius, $s_{\ell m}$ are randomly generated spherical harmonic coefficients, and $Y_\ell^m$ are the orthonormal spherical harmonics. Deviation from perfect spherical geometry is characterized by the relative standard deviation of the radius $\sigma$ and the correlation length of angular change $L_c$ (see Refs. [92–94] for detailed explanation). In this work, the correlation length of angular change is held constant at $L_c = 2\sin\left(\frac{\Gamma}{2}\right)$ with correlation angle set at $\Gamma = 30°$, and the relative standard deviation of the radius is varied as $\sigma = 0.2$ (mild distortion) and $\sigma = 0.8$ (high distortion). The spherical harmonic series defining the radius vector of the Gaussian random particles is truncated at $\ell = 10$, a value deemed sufficient by previous light scattering studies [94,95].

Using this approach, we model systems of two $SiO_2$ particles of the same degree of distortion but unique morphology. Individual particle morphology is maintained with increasing level of distortion by implementing the same Gaussian random variables for a given particle. Distortion can then be amplified by increasing the standard deviation of the radius $\sigma$ for the given set of Gaussian random variables (FIG 6). This approach ensures that particles are of completely irregular shape and allows for direct comparison of conductance as a function of the degree of particle distortion.

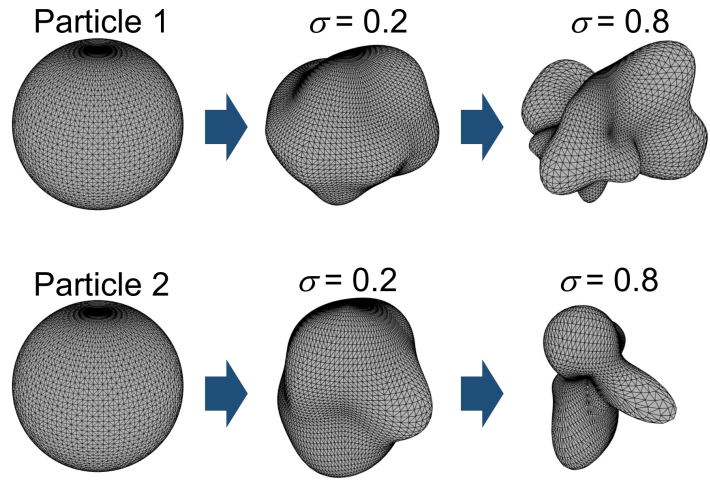

FIG 6. Gaussian random sphere representation of irregularly shaped particles. The relative standard deviation of the radius is varied to define mildly ($\sigma = 0.2$) and highly ($\sigma = 0.8$) distorted particles. Particle morphology is kept constant with increasing particle distortion.

The total conductance between irregularly shaped particles, normalized with respect to the conductance of comparable perfect spheres with equivalent volume and center-of-mass separation distance, is calculated at room temperature ($T = 300$ K) and is shown in FIG 7. Geometric irregularity most strongly affects the total conductance at close separation distances. For center-of-mass separation distances $d_c < 300$ nm, increasing distortion is associated with lower total conductance with respect to that of comparable perfect spheres. At the closest center-of-mass separation distance ($d_c = 110$ nm), the total conductance for mildly and highly distorted particles is reduced to respectively 78% and 64% of the total conductance of comparable perfect spheres. At larger center-of-mass separation distances ($d_c \gtrsim 300$ nm), the total conductance for both mildly and highly distorted particles reaches a plateau of 95% of the total conductance of comparable perfect spheres. Based on the analysis in Sec. III, this 5% difference is close to the numerical error in the DSGF model (< 3%, see FIG 3(a)). As such, we conclude that distortion most significantly affects the total conductance of $R_{eq} \approx 35$–$50$ nm particles when they are closely spaced ($d_c < 300$ nm), and the total conductance of these particles can be approximated by that of perfect spheres at larger center-of-mass separation distances ($d_c \gtrsim 300$ nm).

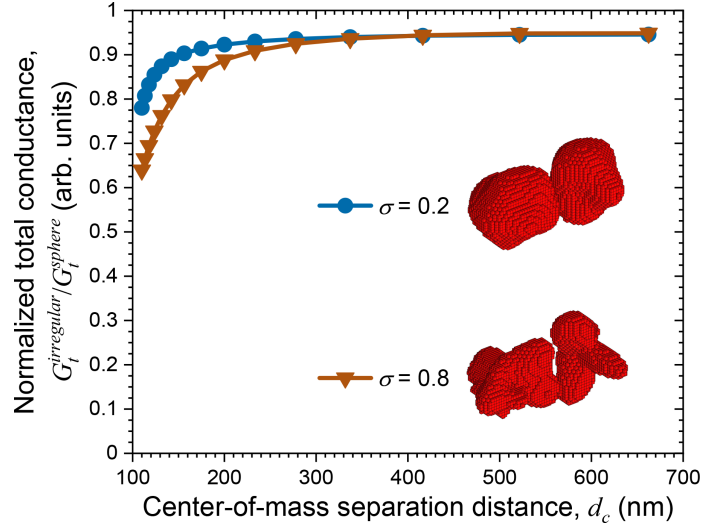

FIG 7. Normalized total conductance as a function of the center-of-mass separation distance for mildly distorted ($\sigma = 0.2$, $R_{eq,1} = 49.3$ nm, $R_{eq,2} = 44.5$ nm, $N = 16{,}605$ total subvolumes) and highly distorted ($\sigma = 0.8$, $R_{eq,1} = 45.8$ nm, $R_{eq,2} = 37.9$ nm, $N = 10{,}948$ total subvolumes) particles, where $R_{eq}$ is the equivalent radius of a perfect sphere with the same volume as the irregularly shaped particle. The total conductance between irregularly shaped particles is normalized with respect to the conductance of comparable perfect spheres with equivalent volume and center-of-mass separation distance.

The underlying physics driving NFRHT between irregularly shaped particles can be revealed through further analysis of the spatial distribution of power dissipation (FIG 8) and the spectral conductance (FIG 9). In the following, we focus on highly distorted particles ($\sigma = 0.8$), though similar, albeit less exaggerated, trends are observed for the mildly distorted particles ($\sigma = 0.2$). We first analyze particles at the closest center-of-mass separation distance $d_c = 110$ nm where particle size is larger than the vacuum separation distance between the closest surfaces $d = 31$ nm. Next, we consider a larger center-of-mass separation distances ($d_c = 337$ nm) for which the normalized total conductance has converged to 95%. For this second case, particle size is smaller than the vacuum separation distance between the closest surfaces $d = 246$ nm.

The spatial distribution of power dissipation illustrates that heat transfer is confined to the nearest surfaces of closely spaced particles (FIG 8(a), $d_c = 110$ nm). In this regime, NFRHT is mostly a surface process dominated by SPhPs with penetration depth $\delta$ approximately equal to the vacuum separation distance ($\delta \approx d = 31$ nm) [89]. From these results, the decreased total conductance seen at small center-of-mass separation distances in FIG 7 can be attributed to the fact that irregularly shaped particles have a smaller volume contributing to NFRHT than do comparable perfect spheres. At a farther center-of-mass separation distance, heat transfer is spread more evenly across the entire volume of particles (FIG 8(b), $d_c = 337$ nm). In this regime, NFRHT is a volumetric process rather than a surface one and is dominated by localized surface phonons (LSPhs) arising from the confinement of SPhPs in subwavelength particles. Therefore, since the volumes of irregularly shaped particles and comparable perfect spheres are the same, the total conductance of irregularly shaped particles at $d_c = 337$ nm is effectively the same as that of comparable perfect spheres, within the margin of error of the DSGF method.

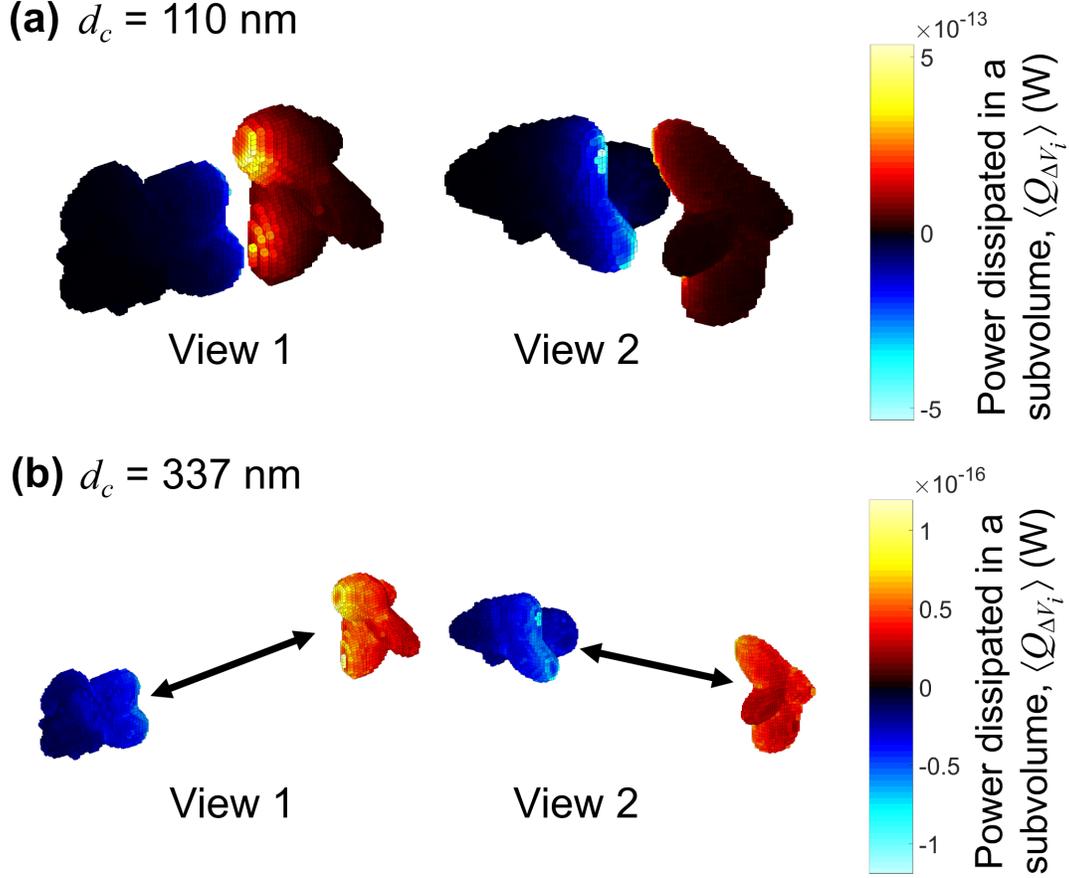

FIG 8. Spatial distribution of power dissipation within each subvolume for two highly distorted particles ($\sigma = 0.8$, $R_{eq,1} = 45.8$ nm, $R_{eq,2} = 37.9$ nm, $N = 10{,}948$ total subvolumes) at center-of-mass separation distances (a) $d_c = 110$ nm and (b) $d_c = 337$ nm (arrows denote which particles are interacting). Negative values represent heat loss, and positive values represent heat gain. The high-temperature particle (blue color scheme since thermal energy is lost) is set at $T = 300$ K, and the low-temperature particle (red color scheme since thermal energy is gained) is set at $T = 0$ K. For $d_c = 110$ nm, 80% of all power dissipation occurs at adjacent surfaces within 34% of the volume of particle 1 and 53% of the volume of particle 2. For $d_c = 337$ nm, 80% of all power dissipation occurs within 67% of the volume of particle 1 and 71% of the volume of particle 2.

The spectral conductance at both small and large center-of-mass separation distances exhibits low- and high-frequency resonances in the Reststrahlen spectral bands of SiO$_2$ ($\sim 8.691 \times 10^{13}$ to $9.656 \times 10^{13}$ rad/s and $\sim 2.038 \times 10^{14}$ to $2.327 \times 10^{14}$ rad/s) where the real component of the dielectric function is negative (FIG 9). For $d_c = 110$ nm, the resonances are due to SPhPs, whereas the resonances for $d_c = 337$ nm are dominated by LSPhs. Across both separation distances, the spectral conductance of irregularly shaped particles displays damping and broadening of resonances (i.e., reduced spectral coherence) as compared with that of comparable perfect spheres. From the full width at half maximum values of each resonance, we estimate that the low- and high-frequency resonances at $d_c = 110$ nm are respectively 57% and 90% broader for irregularly shaped particles than resonances of comparable perfect spheres. The corresponding broadening values for the $d_c = 337$ nm particle system increases to 64% and 174%, respectively. The spectral conductance at other frequencies remains unchanged, because NFRHT at these frequencies is mediated by propagating and frustrated modes (i.e., bulk modes) rather than resonant modes (i.e., SPhPs and LSPhs).

At the closest center-of-mass separation distance of 110 nm where NFRHT is essentially a surface process, SPhP resonances for the case of perfect spheres arise at frequencies where the real component of the dielectric function of $SiO_2$ is approximately equal to -1 (slight deviation from this value occurs due to nonnegligible losses) [15]. For irregularly shaped particles, damping and broadening of resonances can be attributed to coupling of SPhPs within the random distorted features of the particles at adjacent surfaces. This effect is similar to the well-known phenomenon of SPhP coupling in thin films that leads to resonance splitting into symmetric and antisymmetric modes. For NFRHT between thin films, resonance splitting is visible in the spectral conductance when the film thickness is comparable to or smaller than the vacuum separation distance [96,97]. This leads to SPhP resonance broadening and damping compared to the case of thick materials since the resonant frequencies of the symmetric and antisymmetric modes can take any values within the Reststrahlen bands and depend on the film thickness and vacuum separation distance. For the irregularly shaped particles presented here, the distorted features are on the order of or smaller than the vacuum separation distance. As such, SPhPs dominating NFRHT with penetration depth approximately equal to or larger than the length scale of the distorted features may couple within these features and alter the spectrum of conductance, thus resulting in resonance broadening and damping. In addition, owing to the randomness of the irregularly shaped particles, the resonant frequencies of particle 1 are unlikely to be the same as those of particle 2. These non-matching resonances reduce spectral coherence.

The increased resonance broadening at larger center-of-mass separation distances stems from the transition of NFRHT from a SPhP-mediated surface phenomenon to a LSPh-mediated volumetric phenomenon. Since LSPhs arise from the confinement of SPhPs in particles, their resonance frequencies depend strongly on the particle geometry [98,99]. The irregularly shaped particles presented here are defined as Gaussian random spheres, such that the overall particle geometry is the aggregate of many distinct spherical harmonic morphologies (see Sec. S5 of the Supplemental Material [88]). Each of the spherical harmonic morphologies support LSPhs with resonances that depend on the shape. This claim can be better understood by using the electric dipole approximation (in vacuum) to determine LSPh resonance conditions of a few spherical harmonic morphologies.

Power dissipation in an electric dipole is proportional to $\text{Im}(\alpha_i)$, where $\alpha_i$ is the dipole polarizability tensor ($i = x, y, z$) defined as [100]

$$\alpha_i = \frac{4\pi}{3}\varepsilon_0 a_x a_y a_z \frac{\varepsilon(\omega)-1}{1+L_i[\varepsilon(\omega)-1]}. \tag{29}$$

The physical dimensions of the dipole are given by $a_x$, $a_y$ and $a_z$, whereas $L_i$ are factors determined from the dipole geometry (see Refs. [100,101] for details). Since the spherical harmonics are multiplied by randomly generated coefficients $s_{\ell m}$ (see Eq. (28)), the morphologies for particles 1 and 2 are generally different. This implies that, in general, LSPh resonances for particles 1 and 2 are different for a given spherical harmonic morphology $\ell$. The spherical harmonic morphology for $\ell = 0$ is a sphere. For this case, $L_x = L_y = L_z = 1/3$, and LSPh resonance occurs when $|\varepsilon + 2|$ is minimum for both particles 1 and 2. When losses are neglected, this corresponds to $\text{Re}(\varepsilon) = -2$ which is the Fröhlich resonance condition [101]. For $\ell = 1$, the spherical harmonic morphology is a slightly distorted sphere, such that the aforementioned Fröhlich resonance condition can still be used as a reasonable approximation for both particles 1 and 2. For $\ell = 2$, the spherical harmonic morphology can be approximated by an ellipsoid. Here, different resonance conditions are obtained for particles 1 and 2 since their dimensions are different. For particle 1, we estimate that $L_x = 0.195$, $L_y = 0.300$, and $L_z = 0.505$. This results in three distinct LSPh resonances that are satisfied when $|\varepsilon + 4.13|$, $|\varepsilon + 2.33|$, and $|\varepsilon + 0.98|$ are minimum. For particle 2, we estimate that $L_x = 0.494$, $L_y = 0.396$, and $L_z = 0.110$. LSPh resonances for this case arise when $|\varepsilon + 1.02|$, $|\varepsilon + 1.53|$, and $|\varepsilon + 8.09|$ are minimum. We expect that a similar process can be applied to higher order spherical harmonic morphologies. Therefore, when all spherical harmonic morphologies are summed to form one irregularly shaped particle, the final particle should support all the individual LSPhs, thus leading to resonance damping and broadening. Similar damping and broadening effects have been reported in electromagnetic scattering studies of continuous distributions of ellipsoids that support a range of distinct LSPh resonances [100,102]. Finally, in addition to irregularly shaped particles supporting a large number of LSPh modes, we expect that resonance mismatch further reduces spectral coherence between particles 1 and 2 starting at the $\ell = 2$ spherical harmonic morphology.

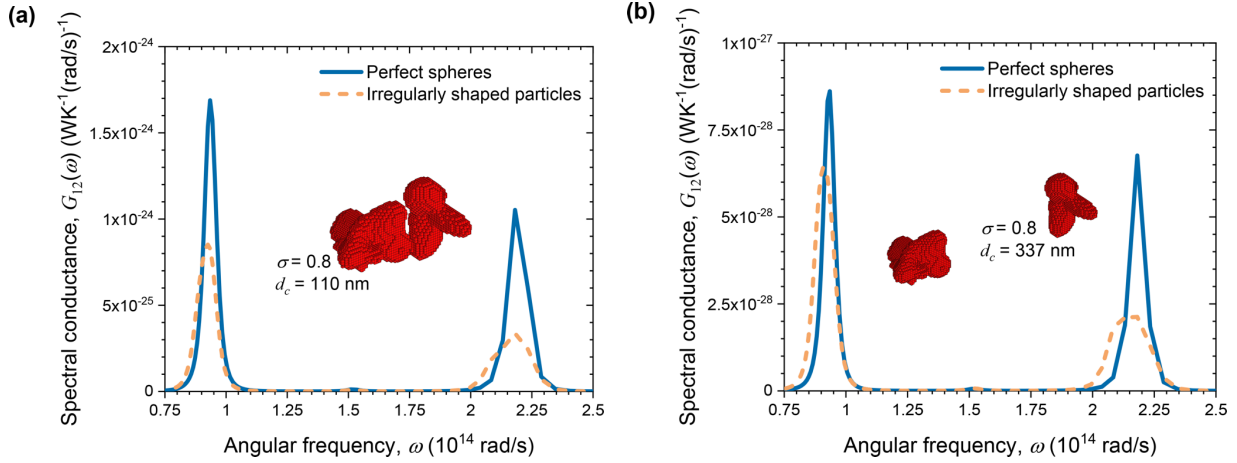

FIG 9. Spectral conductance at $T = 300$ K for two highly distorted particles ($\sigma = 0.8$, $R_{eq,1} = 45.8$ nm, $R_{eq,2} = 37.9$ nm, $N = 10{,}948$ total subvolumes) at center-of-mass separation distances (a) $d_c = 110$ nm and (b) $d_c = 337$ nm. Results are compared against the spectral conductance for two perfect spheres with the same volume and center-of-mass separation distance as the irregularly shaped particles.

## V. CONCLUSIONS

We formulated the DSGF method, which is a volume integral approach based on fluctuational electrodynamics, for predicting NFRHT between finite, three-dimensional objects. The strengths of the DSGF method are that it defines all system interactions independent of the physics of thermal excitation and outputs a general system Green's function parameter to define the system response. We verified the DSGF method against the analytical solution for chains of two and three SiO$_2$ spheres embedded in vacuum. Good agreement was found between the DSGF calculations and analytical solutions. For discretizations in the range of ~2000-6000 subvolumes per sphere, the error in the DSGF-calculated total conductance was below 3% and 5% in cases of two 50-nm-radius spheres and two 500-nm-radius spheres, respectively, at vacuum separation distances in the range $R \lesssim d \lesssim 20R$.

After verification, we applied the DSGF method to study NFRHT in the hitherto unstudied systems of irregularly shaped SiO$_2$ particles, with geometric distortion modeled using the Gaussian random sphere technique. The DSGF results illustrated that, at vacuum separation distances smaller than the particle size (i.e., when NFRHT is essentially a surface phenomenon), irregular geometry led to reduction in the total conductance from that of comparable perfect spheres. We showed that geometric effects are negligible when calculating total conductance values at vacuum separation distances larger than the particle size (i.e., when NFRHT is a volumetric phenomenon). In this regime, even highly distorted particles may be approximated by comparable perfect spheres when the total conductance is desired. For such cases, computationally inexpensive analytical solutions for spheres can be used with little loss of accuracy, and the high computational costs inherent to full-scale numerical models can be avoided.

Particle irregularity resulted in reduced spectral coherence (i.e., broadening and damping of resonances) regardless of the separation distance. When particle size was larger than the vacuum separation distance, the reduced spectral coherence was attributed to coupling of SPhPs within the distorted features of the particles. We showed that reduction of spatial coherence was exacerbated by increasing the separation distance. For the case in which particle size was smaller than the vacuum separation distance, we attribute further resonance broadening to the existence of multiple, distinct LSPh modes supported by the individual spherical harmonic morphologies that compose the Gaussian random spheres. From these spectral analyses, we conclude that it is necessary to use a numerical method like the DSGF to capture geometry-dependent effects and accurately model the spectrum of NFRHT, especially around SPhP and LSPh resonances.

The results presented in this work have important implications for thermal management in micro/nanoscale devices composed of dielectric particles. In particular, these results highlight that geometric defects in real manufactured particles may significantly affect thermal transport and should be modeled rigorously in closely spaced particle designs. Additionally, when the goal is to engineer the spectrum of NFRHT via metamaterials made of particles, it is crucial that full-scale models such as the DSGF be used to account for the impact of geometric irregularities on resonances. As this is the first time in which NFRHT has been studied between irregularly shaped particles, open questions remain. Future research should identify effects of other distortion parameters (such as the correlation length of angular change in the Gaussian random sphere characterization), compare thermal radiation models to real manufactured particle beds [103], and determine more general trends of particle irregularity on thermal radiation by incorporating orientation averaging, although we expect that orientation averaging will lead to qualitatively similar results.

## ACKNOWLEDGMENTS

This work was supported by the National Science Foundation (Grant No. CBET-1952210). L.P.W. acknowledges that this material is based upon work supported by the National Science Foundation Graduate Research Fellowship under Grant No. DGE-1747505. Any opinions, findings, and conclusions or recommendations expressed in this material are those of the authors and do not necessarily reflect the views of the National Science Foundation. L.P.W. is grateful for the helpful discussions with and support received from Dr. Henry Fu in preparation of this manuscript. This work was authored in part by the National Renewable Energy Laboratory (NREL), operated by Alliance for Sustainable Energy, LLC, for the U.S. Department of Energy (DOE) under Contract No. DE-AC36-08GO28308. E.J.T was supported in part by the Laboratory Directed Research and Development (LDRD) Program at NREL. The views expressed in the article do not necessarily represent the views of the DOE or the U.S. Government. The support and resources from the Center for High Performance Computing at the University of Utah are gratefully acknowledged.

# APPENDIX A: HOMOGENEOUS SOLUTION OF THE WAVE EQUATION FOR THE TOTAL ELECTRIC FIELD

The homogeneous solution $\mathbf{E}_0(\mathbf{r}, \omega)$ of Eq. (3) satisfies the wave equation (7). The wave equation (7) may be rearranged into yet another nonhomogeneous wave equation by moving the relative dielectric function, $\varepsilon_r(\mathbf{r}, \omega)$, to the right-hand side

$$\nabla \times \nabla \times \mathbf{E}_0(\mathbf{r}, \omega) - k_0^2 \varepsilon_{ref}(\omega) \mathbf{E}_0(\mathbf{r}, \omega) = i\omega \mu_0 \mathbf{J}^{(eq,0)}(\mathbf{r}, \omega), \mathbf{r} \in \Re^3, \tag{A1}$$

where $\mathbf{J}^{(eq,0)}$ is the equivalent (induced) current density acting as a scattering source due to an external exciting field. $\mathbf{J}^{(eq,0)}$ is defined as [104]

$$\mathbf{J}^{(eq,0)}(\mathbf{r}, \omega) = -i\omega \varepsilon_0 \varepsilon_r(\mathbf{r}, \omega) \mathbf{E}_0(\mathbf{r}, \omega), \mathbf{r} \in \Re^3. \tag{A2}$$

Eq. (A1) may be solved as the summation of the homogeneous and particular solutions,

$$\mathbf{E}_0(\mathbf{r}, \omega) = \mathbf{E}_{0,0}(\mathbf{r}, \omega) + \mathbf{E}_{0,p}(\mathbf{r}, \omega). \tag{A3}$$

The homogeneous solution $\mathbf{E}_{0,0}(\mathbf{r}, \omega)$ of Eq. (A1) satisfies the following wave equation:

$$\nabla \times \nabla \times \mathbf{E}_{0,0}(\mathbf{r}, \omega) - k_0^2 \varepsilon_{ref}(\omega) \mathbf{E}_{0,0}(\mathbf{r}, \omega) = \mathbf{0}, \ \mathbf{r} \in \Re^3. \tag{A4}$$

Since this equation describes the electric field that would exist in the lossless background reference medium without any objects present, the solution is any externally imposed exciting field (i.e., incident field) satisfying Eq. (A4).

The particular solution $\mathbf{E}_{0,p}(\mathbf{r}, \omega)$ of Eq. (A1) is the scattered electric field from the objects illuminated by an external exciting field and may be solved for by using the volume integral technique

$$\mathbf{E}_{0,p}(\mathbf{r}, \omega) = i\omega \mu_0 \int_V \overline{\overline{\mathbf{G}}}^0(\mathbf{r}, \mathbf{r}', \omega) \mathbf{J}^{(eq,0)}(\mathbf{r}', \omega) d^3\mathbf{r}', \ \mathbf{r} \in \Re^3, \tag{A5}$$

where $\overline{\overline{\mathbf{G}}}^0(\mathbf{r}, \mathbf{r}', \omega)$ is the free-space Green's function with known analytical form [76,77] and integration is over all real space. Numerical solution of $\mathbf{E}_0(\mathbf{r}, \omega)$ for scattering objects embedded in a lossless background reference medium may be obtained by the well-established discrete dipole approximation (DDA) method from the light-scattering literature. Open-source DDA programs applicable to three-dimensional arbitrary geometries are widely available (e.g., ADDA [105,106] and DDSCAT [107,108]).

The complete solution to Eq. (3) in expanded form is thus

$$\mathbf{E}(\mathbf{r},\omega) = \mathbf{E}_{0,0}(\mathbf{r},\omega) + i\omega\mu_0 \int_V \bar{\bar{\mathbf{G}}}^0(\mathbf{r},\mathbf{r}',\omega)\, \mathbf{J}^{(eq,0)}(\mathbf{r}',\omega)\, d^3\mathbf{r}' + i\omega\mu_0 \int_V \bar{\bar{\mathbf{G}}}(\mathbf{r},\mathbf{r}',\omega)\mathbf{J}^{(eq)}(\mathbf{r}',\omega)\, d^3\mathbf{r}', \mathbf{r} \in \Re^3. \quad (A6)$$

Here, Eq. (A6) is general, and integration spans all real space. In the case of a lossless background reference medium, integration will be restricted to the domain of the thermal objects, $V_{therm}$.

### APPENDIX B: DISCRETIZED FREE-SPACE GREEN'S FUNCTION

The discretized version of the free-space Green's function is [66,76,77]

$$\bar{\bar{\mathbf{G}}}^0(\mathbf{r}_i,\mathbf{r}_j,\omega) = \frac{\exp\left(ik_0\sqrt{\varepsilon_{ref}(\omega)}r_{ij}\right)}{4\pi r_{ij}} \left[ \left(1 - \frac{1}{\varepsilon_{ref}(\omega)(k_0 r_{ij})^2} + \frac{i}{k_0\sqrt{\varepsilon_{ref}(\omega)}r_{ij}}\right)\bar{\bar{\mathbf{I}}} - \left(1 - \frac{3}{\varepsilon_{ref}(\omega)(k_0 r_{ij})^2} + \frac{3i}{k_0\sqrt{\varepsilon_{ref}(\omega)}r_{ij}}\right)\left(\hat{\mathbf{r}}_{ij}\hat{\mathbf{r}}_{ij}^\dagger\right) \right] \text{ for } j \neq i, \quad (B1)$$

where $r_{ij} = |\mathbf{r}_i - \mathbf{r}_j|$, $\hat{\mathbf{r}}_{ij} = \frac{(\mathbf{r}_i - \mathbf{r}_j)}{|\mathbf{r}_i - \mathbf{r}_j|}$ and † signifies the conjugate transpose. At the point $\mathbf{r}_i = \mathbf{r}_j$, the free-space Green's function has a singularity. This singularity is circumvented by employing the principal value technique as presented by van Bladel [83] and Yaghjian [84]. For a cubic mesh, the principal value solution of the singularity point is

$$\bar{\bar{\mathbf{G}}}^0(\mathbf{r}_i,\mathbf{r}_j,\omega) = \frac{\bar{\bar{\mathbf{I}}}}{3\Delta V_j \varepsilon_{ref}(\omega) k_0^2}\left\{2\left[e^{ia_j k_0\sqrt{\varepsilon_{ref}(\omega)}}\left(1 - ia_j k_0\sqrt{\varepsilon_{ref}(\omega)}\right) - 1\right] - 1\right\} \text{ for } j = i, \quad (B2)$$

where $a_j$ is the equivalent radius of a subvolume defined as $a_j = \left(\frac{3\Delta V_j}{4\pi}\right)^{\frac{1}{3}}$.

# APPENDIX C: COMPARISON OF THE DSGF AND TDDA METHODS

In both the TDDA and DSGF methods, power dissipation is proportional to the unknown quantity $\bar{\bar{X}}$ that satisfies a matrix equation of the form

$$\bar{\bar{A}}\bar{\bar{X}}\bar{\bar{A}}^\dagger = \bar{\bar{B}}, \tag{C1}$$

where $\bar{\bar{A}}$ and $\bar{\bar{B}}$ are known matrices. In the TDDA, $\bar{\bar{A}}$ is the interaction matrix, $\bar{\bar{X}}$ is the autocorrelation of the total dipole moments $\langle \bar{P}^{(\text{total})} \bar{P}^{(\text{total})\dagger} \rangle$, and $\bar{\bar{B}}$ is the diagonal matrix representing the autocorrelation of thermally fluctuating dipole moments $\langle \bar{P}^{(\text{fl})} \bar{P}^{(\text{fl})\dagger} \rangle$. In this way, Eq. (C1) in the TDDA is written as

$$\bar{\bar{A}} \langle \bar{P}^{(\text{total})} \bar{P}^{(\text{total})\dagger} \rangle \bar{\bar{A}}^\dagger = \langle \bar{P}^{(\text{fl})} \bar{P}^{(\text{fl})\dagger} \rangle. \tag{C2}$$

Mathematically, Eq. (C2) is the eigendecomposition of the Hermitian matrix $\langle \bar{P}^{(\text{total})} \bar{P}^{(\text{total})\dagger} \rangle$ into its real eigenvalues $\langle \bar{P}^{(\text{fl})} \bar{P}^{(\text{fl})\dagger} \rangle$ and orthogonal eigenvectors $\bar{\bar{A}}^{-1}$.

In the DSGF method, $\bar{\bar{A}}$ in Eq. (C1) is the same interaction matrix as in the TDDA, the unknown $\bar{\bar{X}}$ is the outer product of the system Green's function $\bar{\bar{G}}\bar{\bar{G}}^\dagger$, and $\bar{\bar{B}}$ is the outer product of the free-space Green's function $\bar{\bar{G}}^0 \bar{\bar{G}}^{0\dagger}$. Eq. (C1) in the DSGF method becomes

$$\bar{\bar{A}}\bar{\bar{G}}\bar{\bar{G}}^\dagger \bar{\bar{A}}^\dagger = \bar{\bar{G}}^0 \bar{\bar{G}}^{0\dagger}. \tag{C3}$$

In the TDDA approach, Eq. (C2) must be solved directly for the unknown matrix $\langle \bar{P}^{(\text{total})} \bar{P}^{(\text{total})\dagger} \rangle$. The DSGF approach, on the other hand, allows alternative solution methods for the unknown $\bar{\bar{G}}\bar{\bar{G}}^\dagger$ matrix by first solving the simpler equation

$$\bar{\bar{A}}\bar{\bar{G}} = \bar{\bar{G}}^0, \tag{C4}$$

where the system Green's function is the unknown. The TDDA cannot be simplified in this manner because of its reliance on autocorrelation functions of stochastic dipole moments in Eq. (C2).

Setting up the system of equations as Eq. (C4) provides the DSGF approach with two advantages over the TDDA. First, the system Green's function of the DSGF approach is a more general parameter than the autocorrelation of total dipole moments found in the TDDA. By solving directly for the system Green's function in Eq. (C4), the DSGF method outputs the general electromagnetic response of the system to any induced source, whether it be generated thermally or otherwise. In this way, the solution to the main system of equations in the DSGF approach may be post-processed to solve for a variety of quantities of interest. Conversely, the autocorrelation of total dipole moments in the TDDA is a decisively thermal quantity. The second advantage of the DSGF approach is that the DSGF matrix equation provided in Eq. (C4) is of the familiar $\bar{\bar{\mathbf{A}}}\bar{\bar{\mathbf{X}}} = \bar{\bar{\mathbf{B}}}$ form. This matrix form has more known solution algorithms than the matrix equations (C1)–(C3) of form $\bar{\bar{\mathbf{A}}}\bar{\bar{\mathbf{X}}}\bar{\bar{\mathbf{A}}}^\dagger = \bar{\bar{\mathbf{B}}}$. As such, well-known algorithms may be applied directly in the DSGF approach to reduce computational loads.

# Supplemental Material

# Near-field radiative heat transfer between irregularly shaped dielectric particles modeled with the discrete system Green's function method

Lindsay P. Walter[1], Eric J. Tervo[2,3,*], and Mathieu Francoeur[1,†]

[1]Radiative Energy Transfer Lab, Department of Mechanical Engineering, University of Utah, Salt Lake City, UT 84112, USA

[2]Department of Electrical and Computer Engineering, University of Wisconsin-Madison, Madison, WI 53706, USA

[3]Department of Mechanical Engineering, University of Wisconsin-Madison, Madison, WI 53706, USA

* Corresponding author. Email address: tervo@wisc.edu
† Corresponding author. Email address: mfrancoeur@mech.utah.edu

## S1. DIELECTRIC FUNCTION OF SiO$_2$

The dielectric function of SiO$_2$ used in the simulations is taken from Ref. [1] and is shown in FIG S1.

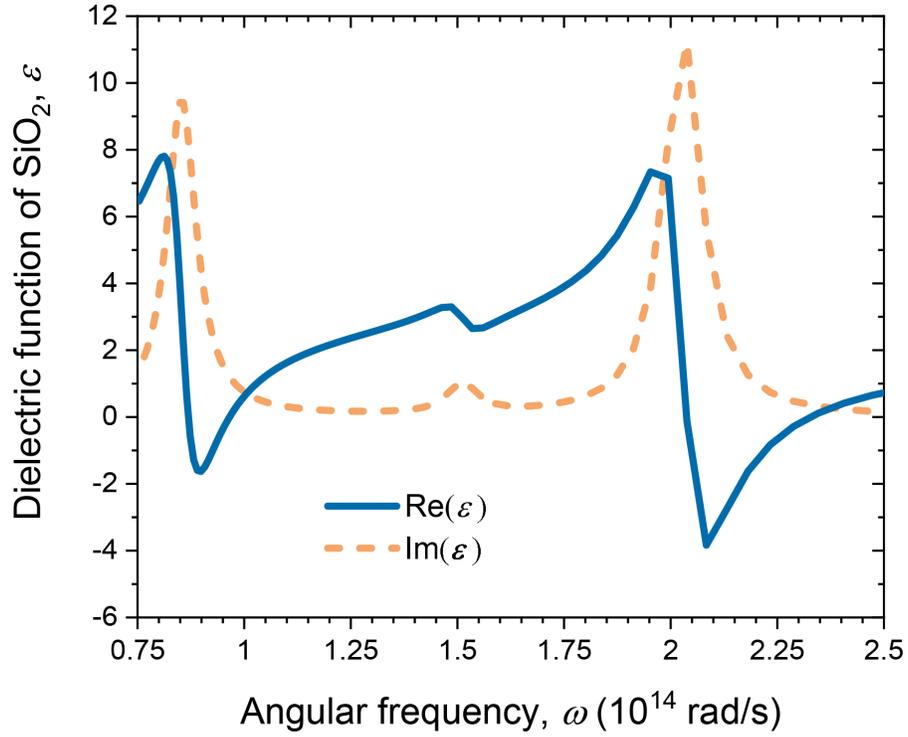

FIG S1. Dielectric function of SiO$_2$ defined from the Lorentz oscillator model with parameters given from Ref. [1].

# S2. EQUIVALENCE OF DISCRETE SYSTEM GREEN'S FUNCTION (DSGF) AND THERMAL DISCRETE DIPOLE APPROXIMATION (TDDA) RESULTS

We compare spectral conductance at $T = 400$ K generated using the discrete system Green's function (DSGF) method against published results calculated with the thermal discrete dipole approximation (TDDA) [2]. The system consists of two 250-nm-radius $SiO_2$ spheres embedded in vacuum. The vacuum separation distance is 200 nm. For the same discretization of 552 subvolumes per sphere, the DSGF approach yielded equivalent results for the spectral conductance as the TDDA (FIG S2). The negligibly small relative difference between DSGF- and TDDA-calculated spectral conductance is on the order of $O(10^{-12})$% and is attributed to computational round-off error.

The equivalence of the TDDA and DSGF approaches in modeling near-field radiative heat transfer is expected because both methods are derived from the same principles of fluctuational electrodynamics and share the same numerical approximations to convert from a continuous to a discretized system. Additional analyses were conducted for variable discretization refinement (not shown), and convergence behavior of the DSGF method was found equivalent to that of the TDDA. From this assessment, it is concluded that published error and convergence evaluations for the TDDA [3] are also applicable to the DSGF method.

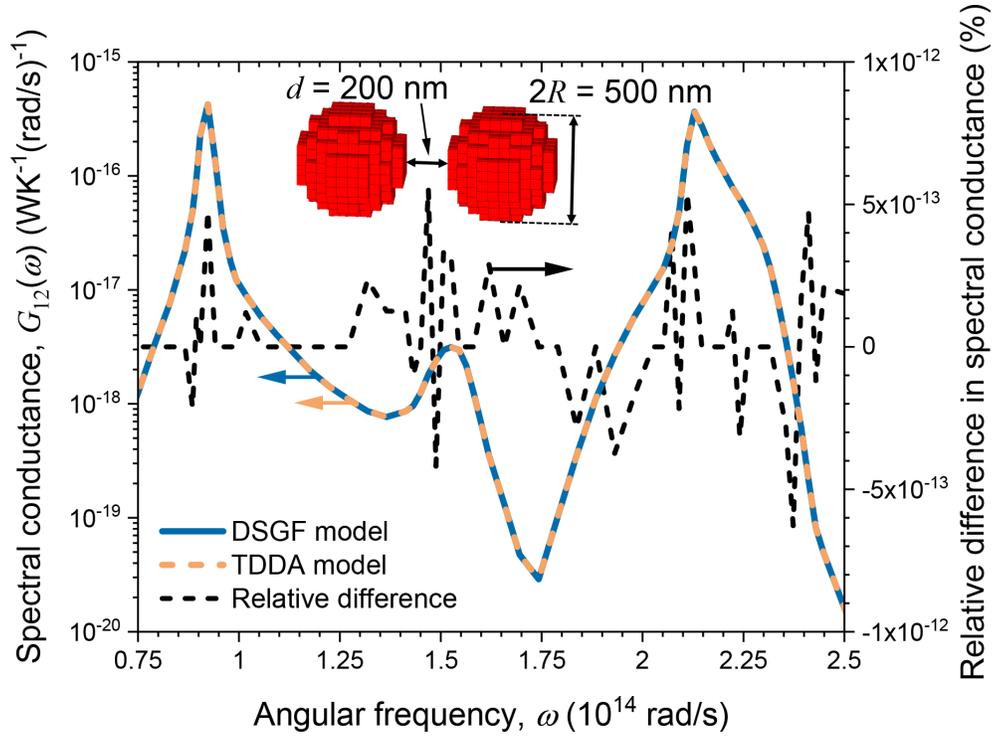

FIG S2. Comparison of the spectral conductance at $T = 400$ K between two $SiO_2$ spheres calculated by the TDDA [2] and by the DSGF method. Spheres of radius $R = 250$ nm are modeled for a vacuum separation distance $d = 200$ nm. The spheres are discretized into $N_{sphere} = 552$ subvolumes per sphere. The relative difference between the TDDA and DSGF is calculated as $\frac{G_{12}^{DSGF}(\omega) - G_{12}^{TDDA}(\omega)}{G_{12}^{TDDA}(\omega)}$.

## S3. SPECTRAL CONDUCTANCE BETWEEN TWO SiO$_2$ SPHERES

The spectral conductance between two 50-nm-radius spheres and two 500-nm-radius spheres at vacuum separation distances $d \approx R$ and $d \approx 20R$ is plotted in FIG S3. These vacuum separation distances are the small and large vacuum separation distance limits for which the DSGF method produced total conductance calculations with minimal numerical error. For all cases, the DSGF calculations capture well the trends in the spectral distribution of conductance. Good agreement with the analytical solution is obtained at the surface phonon-polariton and localized surface phonon resonances that dominate heat transfer. Surface phonon-polariton resonances are found in the regimes presented here where $d \approx R$ (FIG S3(a) and FIG S3(c)), and localized surface phonons are found in the regimes presented here where $d \approx 20R$ (FIG S3(b) and FIG S3(d)).

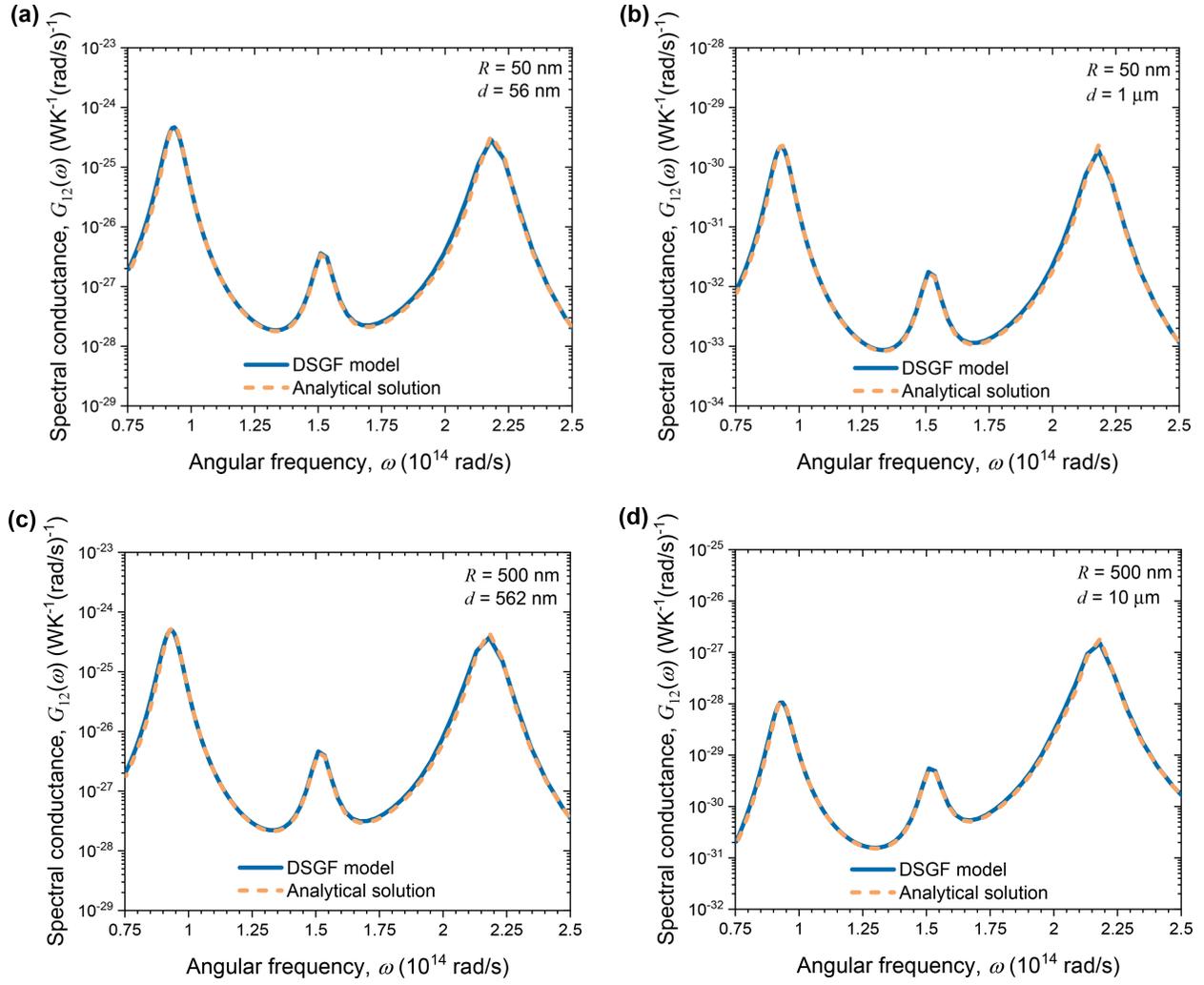

FIG S3. Comparison of the spectral conductance at $T = 300$ K between two SiO$_2$ spheres calculated analytically [1] and by the DSGF method when $N_{sphere} = 2176$ subvolumes per sphere are used in the discretization. (a) $R = 50$ nm and $d = 56$ nm. (b) $R = 50$ nm and $d = 1$ μm. (c) $R = 500$ nm and $d = 562$ nm. (d) $R = 500$ nm and $d = 10$ μm.

# S4. DISCRETE SYSTEM GREEN'S FUNCTION (DSGF) RESULTS FOR THREE SiO$_2$ SPHERES CALCULATED WITH ONE SUBVOLUME PER SPHERE

The total conductance for a linear chain of three 50-nm-radius SiO$_2$ spheres is shown in FIG S4. The spheres are discretized into one subvolume per sphere. The total conductance between the first and second sphere and between the outer spheres is reduced below 1% for vacuum separation distances $d > 20R$ ($d > 1000$ nm).

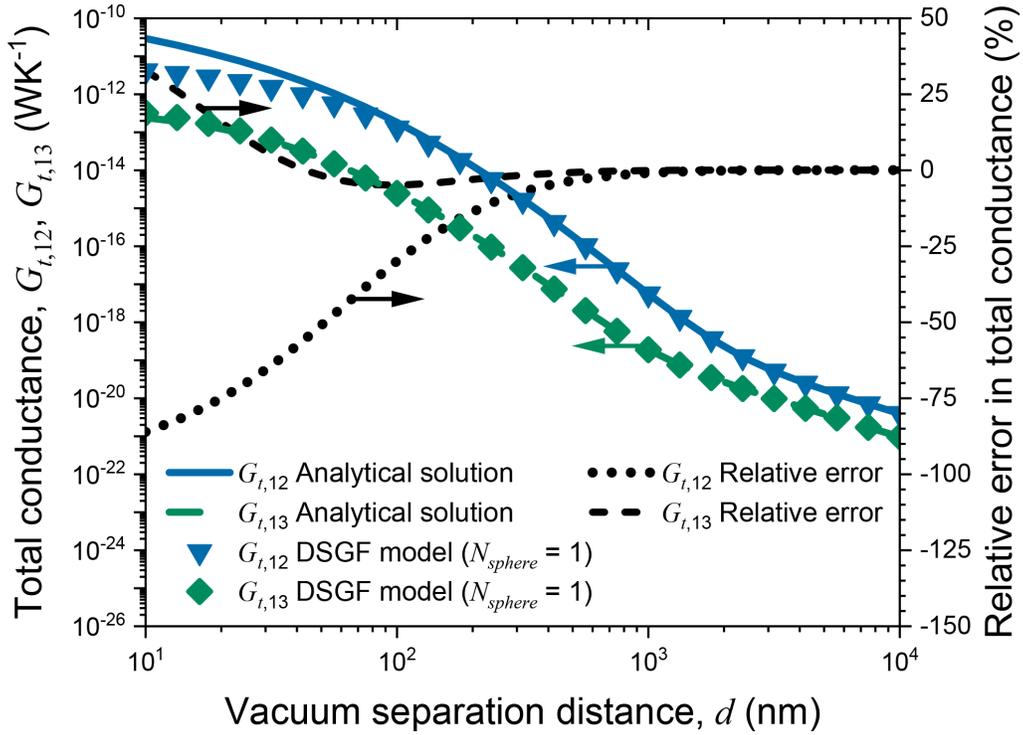

FIG S4. Comparison of the total conductance at $T = 300$ K between three 50-nm-radius SiO$_2$ spheres calculated analytically [1] and by the DSGF method when one subvolume per sphere is used in the discretization. Vacuum separation distance is varied as 10 nm $\leq d \leq$ 10 μm. The relative error between the DSGF and analytical solution is calculated as $\frac{G_{t,12}^{DSGF} - G_{t,12}^{analytical}}{G_{t,12}^{analytical}}$ and $\frac{G_{t,13}^{DSGF} - G_{t,13}^{analytical}}{G_{t,13}^{analytical}}$.

# S5. SPHERICAL HARMONIC DECOMPOSITIONS OF GAUSSIAN RANDOM SPHERES

The Gaussian random spheres defining particles 1 and 2 are decomposed into shapes defined from spherical harmonics of a given degree $\ell$ (see FIG S5 for decomposition of particle 1 and FIG S6 for decomposition of particle 2). The logarithmic radius of each decomposed shape is defined as $s(\theta, \varphi, \ell) = \sum_{m=-\ell}^{\ell} s_{\ell m} Y_\ell^m(\theta, \varphi)$, where $\theta$ is the polar angle, $\varphi$ is the azimuthal angle, $s_{\ell m}$ are randomly generated spherical harmonic coefficients, and $Y_\ell^m$ are the orthonormal spherical harmonics.

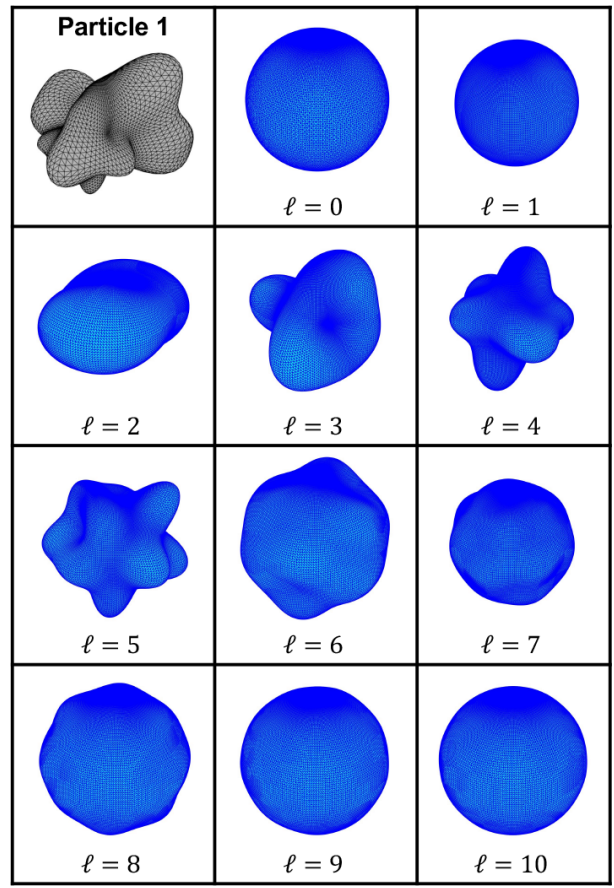

FIG S5. Decomposition of particle 1 into shapes defined from spherical harmonics of degree $0 \leq \ell \leq 10$.

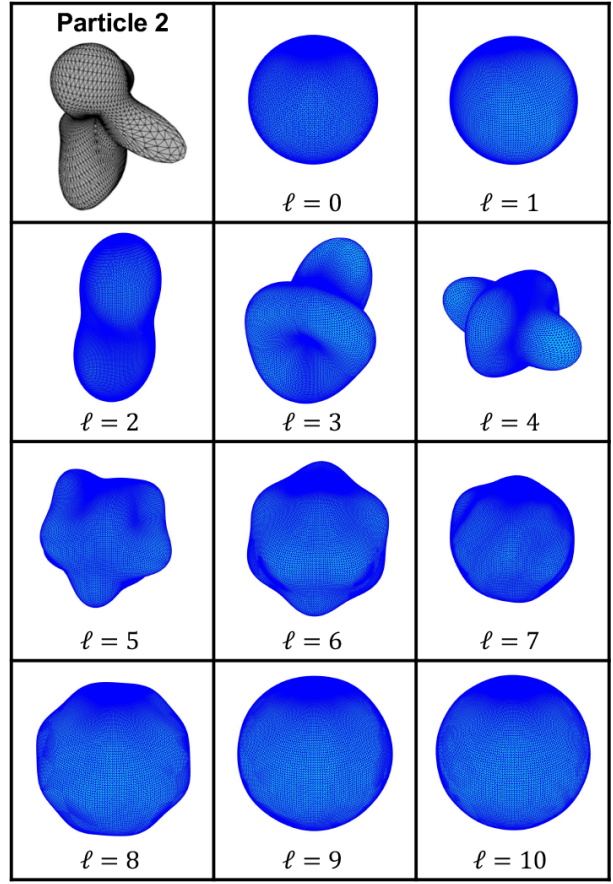

FIG S6. Decomposition of particle 2 into shapes defined from spherical harmonics of degree $0 \leq \ell \leq 10$.